\documentclass[aps,prd,twocolumn,tightenlines,showpacs,amsmath,amssymb,nofootinbib,superscriptaddress]{revtex4-1}

\usepackage{latexsym}
\usepackage{graphicx}
\usepackage{subfigure}
\usepackage{cases}
\usepackage{hyperref}
\usepackage{amssymb}
\usepackage{bm}
\usepackage{natbib}
\usepackage{amsmath}
\usepackage{slashed}
\usepackage{braket}
\usepackage{color}  

\usepackage{bm}
\usepackage{tensor}
\usepackage{hyperref}
\usepackage{slashed}
\usepackage{multirow}
\usepackage{natbib}
\usepackage{braket} % \Bra{\psi} , \Ket {\psi} , \Braket{\psi|\psi}

\newcommand{\be}{\begin{equation}}
\newcommand{\ee}{\end{equation}}
\newcommand{\ba}{\begin{eqnarray}}
\newcommand{\ea}{\end{eqnarray}}
\def\bal{\begin{align}}
\def\eal{\end{align}}
\def \bse {\begin{subequations} \begin{eqnarray}}
\def \ese {\end{eqnarray} \end{subequations}}
\def\bald{\begin{aligned}}
\def\eald{\end{aligned}}

\newcommand{\per}{\, .}
\newcommand{\com}{\, ,}
\newcommand{\eref}[1]{Eq.~(\ref{#1})}

\def\L{{\scriptscriptstyle L}}
\def\R{{\scriptscriptstyle R}}

\def\bmx{{\bm x}}
\def\bmp{{\bm p}}
\def\bmk{{\bm k}}
\def\lb{\left[}
\def\rb{\right]}
\def\inp{\int \dbar^3p}

\def\inx{\int d^3x}
\newcommand{\dbar}{\mathchar'26\mkern-9mu d}

\begin{document}

\title{Chiral Gravitational Waves from Chiral Fermions}

\date{\today}

\author{Mohamed M. Anber}
\email[]{manber@lclark.edu}
\affiliation{Department of Physics, Lewis $\&$ Clark College, 
Portland, OR 97219, USA}
\affiliation{Laboratory of Particle Physics and Cosmology, 
Institute of Physics, \\
Ecole Polytechnique F\'ed\'erale de Lausanne
CH-1015 Lausanne
Switzerland}
\author{Eray Sabancilar}
\email[]{eray.sabancilar@physik.uni-bielefeld.de}
\affiliation{Laboratory of Particle Physics and Cosmology, 
Institute of Physics, \\
Ecole Polytechnique F\'ed\'erale de Lausanne
CH-1015 Lausanne
Switzerland}
\affiliation{Department of Physics, Bielefeld University, D-33615, Bielefeld, Germany}

%========================================================================================
\begin{abstract}
We report on a new mechanism that leads to the generation of primordial chiral gravitational waves, and hence, the violation of the parity symmetry in the Universe. We show that nonperturbative production of fermions with a definite helicity is accompanied by the generation of chiral gravitational waves. This is a generic and model-independent phenomenon that can occur during inflation, reheating and radiation eras, and can leave imprints in the cosmic microwave background polarization and may be observed in future ground- and space-based  interferometers. We also discuss a specific model where chiral gravitational waves are generated via the production of light chiral fermions during pseudoscalar inflation.
\end{abstract}

%=================================================================================
\maketitle

%=======================================================================================
\section{Introduction}
\label{sec:introduction}
%=======================================================================================

Gravitational waves (GW) are an invaluable probe for studying the early Universe as well as various astrophysical events. Recently,  there has been a tremendous effort to detect them using both direct and indirect methods. Primordial gravitational waves, i.e., GW of cosmological origin, leave imprints on the polarization of the cosmic microwave background (CMB). There are several ongoing and upcoming experiments dedicated to observe the CMB polarization signal, such as Keck Array \cite{Array:2015xqh}, BICEP3 \cite{Ahmed:2014ixy}, PolarBEAR \cite{Ade:2014afa}, SPTpol \cite{Keisler:2015hfa} and ACTpol \cite{Niemack:2010wz}. In addition, the recent detection of GW from the astrophysical events GW150914 and GW151226 by LIGO-Virgo collaboration \cite{Abbott:2016blz,Abbott:2016nmj} has inspired the interest in the direct detection of the cosmological GW in ground and space-based interferometers. In fact, primordial GW will open a new observational window into the very first moments of the Universe.

One of the important questions about the Universe is whether parity ($\cal P$) is respected or broken on macroscopic scales\footnote{Parity violation has a paramount importance in explaining the baryon asymmetry of the Universe. A possible connection between various baryogenesis mechanisms and macroscopic violation of $\cal P$ have been considered in Refs.~\cite{Joyce:1997uy,Brustein:1998du,Giovannini:1997eg,Giovannini:1999wv,Giovannini:1999by,Bamba:2006km,Bamba:2007hf,Fujita:2016igl,Domcke:2016bkh,Ben-Dayan:2016iks,Adshead:2016iae,Kamada:2016eeb,Anber:2015yca,Alexander:2016moy,Alexander:2004us,Adshead:2015jza,Long:2013tha,Sabancilar:2013raa}.
}. 
Observation of chiral configurations of electromagnetic or gravitational fields on macroscopic scales will be a strong evidence for $\cal P$ violation. 
Chiral (circularly polarized) GW of primordial origin can give rise to non-zero TB and EB cross correlators in the CMB \cite{Gluscevic:2010vv}. Furthermore, they can also be directly detected in an array of ground and space-based interferometers \cite{Hayama:2016kmv,Aasi:2014erp,Cook:2011hg}.

Only a few mechanisms that can generate primordial chiral GW in the early Universe have been proposed in the literature. The first mechanism  involves the coupling of a time varying scalar field to the gravitational topological term, i.e., $\Delta {\cal L} \propto \varphi R_{\mu\nu\rho\sigma} {\tilde R}^{\mu\nu\rho\sigma}$ \cite{Choi:1999zy,Lue:1998mq,Lyth:2005jf,Satoh:2007gn}. The other mechanism that produces chiral GW relies on the generation of helical gauge field configurations \cite{Garretson:1992vt,Anber:2006xt,Caprini:2014mja} via the term $\Delta {\cal L} \propto \varphi F_{\mu\nu} {\tilde F}^{\mu\nu}$ \cite{Sorbo:2011rz,Anber:2012du,Adshead:2013qp,Adshead:2013nka,Maleknejad:2014wsa,Maleknejad:2016qjz,Obata:2016tmo,Obata:2016xcr}. These helical gauge fields, in turn, contribute to the anisotropic stress tensor and source chiral GW. Also, it has been pointed out that chiral GW are produced in Horava-Lifshitz gravity \cite{Takahashi:2009wc}.

In this work, we report on a new model-independent mechanism that generates primordial chiral GW. We show that production of light chiral fermions in a time varying background is accompanied by chiral GW. Light chiral fermions with mass much smaller than the Hubble scale, $m/H \ll 1$, have definite helicity since the helicity flip process is negligible as it is suppressed by $(m/H)^2$. Production of fermions with a definite helicity leads to an asymmetry between the two components of the energy momentum tensor projected along the helicity eigenbasis. Hence, an imbalance between left and right helicities of GW will be created. Therefore, any mechanism that creates an asymmetry between light left and right-chiral fermions will also lead to chiral GW.

This paper is organized as follows.
In Sec.~\ref{sec:notation}, we start by developing the necessary formalism to study the generation of gravitational waves via the production of left-chiral Weyl fermions in the Fridmann-Robertson-Walker (FRW) background. The formalism used to study the generation of GW from fermions is not new, see, e.g., \cite{Enqvist:2012im, Barnaby:2012xt,Figueroa:2014aya,Figueroa:2013vif}. The new component in this work, however, is the generation of helical GW due to the presence of chiral fermions, which is shown by computing the two point functions of the tensor perturbations in the helicity eigenbasis. At the end of Sec.~\ref{sec:notation}, we show that there is an imbalance between the correlation functions of the left and right-helical components of GW. Our main results are given by \eref{h correlator final} and \eref{right h correlator} for left and right-chiral fermions respectively.
In Sec.~\ref{sec:inflation}, we give an explicit example where chiral GW are generated from the production of chiral fermions during pseudoscalar inflation. 
We conclude with a brief summary and discussion of our results in Sec.~\ref{sec:discussion}.

%=======================================================================================
\section{Theory and Formulation}
\label{sec:notation}
%=======================================================================================
In this section, we will study the gravitational wave generation due to the production of left-chiral Weyl fermions from vacuum in a time varying curved background. By left-chiral Weyl fermions we mean  fields with spin $1/2$ that transform in the irreducible representation  $(\frac{1}{2},0)$ of the Lorentz group. In addition to chirality, one can also talk about helicity which is the projection of the spin  $\bm S$ of a particle along its momentum $\bmp$, i.e. we define the helicity as the eigenvalue of the operator $\hat h\equiv 2\bm S\cdot \bmp/p$.  A massive left-chiral Weyl fermion can have both helicities $\pm 1$. In general, helicity is not a Lorentz-invariant quantity since different observers measure different values of $h$ depending on the relative velocity between the fermion and the observer. However, in the mass goes to zero limit we find that the helicity $h=+1$ decouples. Or in other words, a massless left-chiral Weyl fermion has a definite helicity $h=-1$, and thus, helicity and chirality coincide in the massless limit and both serve as a good Lorentz invariant quantum number.

In the following analysis we assume that the mass of the left-chiral Weyl fermion is very small compared to the Hubble's parameter, and hence, we set the mass to zero. We show, in a model-independent way, that the production of chiral fermions from vacuum will always accompany the generation of chiral gravitational waves.

%===========================================
\subsection{Chiral Gravitational Waves}
%===========================================

In this part we review the general formalism used to study the gravitational wave production due to the presence of anisotropic energy-momentum tensor in the Friedmann-Robertson-Walker (FRW) background. In particular, we are interested in studying the two-point functions of the tensor perturbations projected along the helicity eigenbasis.  To this end, we use natural units $c=1$, $\hbar = 1$ and start with the FRW metric written in the conformal coordinates as $ds^2 = a(\tau)^2 (d\tau^2 - d{\bm x}^2)$, where  the conformal time $\tau$ is related to the cosmic time $t$ via $dt=a d\tau$. In what follows derivatives with respect to the cosmic time will be denoted by $d/dt \equiv~ \dot~$, while derivatives with respect to the conformal time will be denoted by $d/d\tau \equiv~ '$. The Hubble parameter is given by $H \equiv \dot a/a$ and the conformal Hubble parameter is ${\cal H} \equiv a'/a = a H$. 

 Introducing the tensor perturbations $h_{\mu\nu}$ to the FRW background metric, we have
\begin{eqnarray}
ds^2=a^2(\tau)\left[\eta_{\mu\nu}+h_{\mu\nu}\right]dx^\mu dx^\nu\,,
\end{eqnarray}
where $\eta_{\mu\nu}=\mbox{diag}\left(1,-1,-1,-1\right)$.
Then, the linearized equation of motion for the tensor perturbations $h_{\mu\nu} (x, \tau)$ is given by
\be
\lb \partial_\tau^2+ 2{\cal H} \partial_\tau - {\bm \nabla}^2  \rb h_{\mu\nu}(\bmx,\tau) = \frac{2}{m_p^2}~ T_{\mu\nu} (\bmx, \tau) \com
\ee
where $T_{\mu\nu} (\bmx, \tau)$ is the physical energy momentum tensor and $m_p$ is the reduced Planck mass. One can also take the Fourier transform of the above equation to obtain
\be\label{h k}
\lb \partial_\tau^2 +2{\cal H} \partial_\tau +k^2  \rb h_{\mu\nu}(\bmk,\tau) = \frac{2}{m_p^2}~ T_{\mu\nu} (\bmk, \tau) \com
\ee
where $k \equiv |\bmk|$ is the comoving momentum. 

Gravitational waves have two physical degrees of freedom that can be expressed using the polarization tensor $\epsilon_{\lambda}^{\mu\nu}(\bmk)$ with appropriate gauge fixing conditions. In what follows we will be interested in the circular helicity modes of gravitational waves, and thus, we use the following parametrization 
\be\label{pol tensor}
\epsilon_{\lambda}^{\mu\nu}(\bmk) = \epsilon_{\lambda}^{\mu}(\bmk) \epsilon_{\lambda}^{\nu}(\bmk) \com
\ee
and we define the circular polarization 4-vectors $\epsilon_{\lambda}^{\mu}(\bmk)$ as
\be
\epsilon_{\lambda}^{\mu}(\bmk)  = \frac{1}{\sqrt{2}} \left[ 0, {\bm \epsilon}_1(\bmk)+ i \lambda {\bm \epsilon}_2(\bmk) \right]\,.
\ee
The set $\left\{\bm \epsilon_1(\bmk), \bm \epsilon_2(\bmk), \hat \bmk \right\}$ form an orthonormal basis, where $\hat \bmk \equiv \bmk / k=\hat {\bm e}_3$, and $\hat {\bm e}_3$ is a unit vector along the third axis.  The polarization takes two values $\pm1$ which we simply denote as $\lambda = \pm$. One can easily check that the following relations are satisfied
\ba
\bald
&\epsilon_{\lambda}^{\mu~*} (\bmk) = \epsilon_{-\lambda}^{\mu}(\bmk)\,, \quad
\epsilon_{\lambda}^{\mu}(-\bmk) = - \epsilon_{-\lambda}^{\mu}(\bmk) \,,\quad
\\
&k \cdot \epsilon_{\lambda}(\bmk) = 0\,,\quad
\epsilon_{\lambda}(\bmk) \cdot \epsilon_{\lambda'}^* (\bmk) = -\delta_{\lambda, \lambda'} \com
\\
&\epsilon_{\lambda}(\bmk) \cdot \epsilon_{\lambda'} (\bmk) =-\delta_{\lambda,- \lambda'}  
\per
\eald
\ea
Also, notice that the  polarization tensor (\ref{pol tensor}) is consistent with the gauge fixing conditions $h_{00}=h_{0i}=h_i^i=h_{i,j}^i=0$, where $i,j$ denote the spatial coordinates. 

Using \eref{pol tensor},  the gravitational perturbations $h^{\mu\nu}$ can be expressed in terms of the circular helicity modes $h_{\pm}$ as
\be\label{h circular}
h^{\mu\nu} (\bmk, \tau) = \sum_{\lambda = \pm} \epsilon_{\lambda}^{\mu}(\bmk) \epsilon_{\lambda}^{\nu}(\bmk) h_{\lambda} (\bmk, \tau) \per
\ee
The inverse relation can be obtained upon contracting \eref{h circular} with the polarization tensor $\epsilon_{-\lambda'}^{\mu\nu}(\bmk)$, which yields
\ba
\bald
\epsilon_{-\lambda'}^{\mu\nu}(\bmk) h_{\mu\nu}(\bmk,\tau) &= \sum_{\lambda = \pm} \epsilon_{\lambda}^{\mu}(\bmk) \epsilon_{\lambda}^{\nu}(\bmk)  \epsilon_{-\lambda'}^{\mu}(\bmk) \epsilon_{-\lambda'}^{\nu}(\bmk) h_{\lambda} (\bmk, \tau)  
\\
&= h_{\lambda'} (\bmk, \tau) \per
\eald
\ea
The decomposition of $h_{\mu\nu}(\bmk,\tau)$ in \eref{h circular} satisfies the transverse traceless conditions $h_{00}=h_{0i}=h_i^i=h_{i,j}^i=0$, and hence, this decomposition projects only the physical degrees of freedom. This can be checked by contracting $h_{\mu\nu}$ with the transverse traceless projection tensor 
\ba
\bald
\Pi_{\mu\nu,\alpha\beta} (\bmk) &= {\cal P}_{\mu\alpha} (\bmk) {\cal P}_{\nu\beta} (\bmk) - \frac{1}{2}{\cal P}_{\mu\nu} (\bmk) {\cal P}_{\alpha\beta} (\bmk)
\com
\\
{\cal P}_{\mu\nu}(\bmk) &\equiv \delta_{\mu\nu} - \frac{k_\mu k_\nu}{k^2} \per
\eald
\ea
Now, the polarization tensor satisfies the following identity:
\be
\epsilon_{\lambda}^{\mu\nu}(\bmk) ~\Pi_{\mu\nu}^{~~~\alpha\beta} (\bmk) = \epsilon_{\lambda}^{\alpha\beta}(\bmk) \per
\ee
Thus, operating with $\Pi_{\mu\nu,\alpha\beta} (\bmk)$ and then with $\epsilon_{-\lambda'}^{\mu\nu}(\bmk)$ on \eref{h k} we obtain the equation of motion for the circular helicity modes of the gravitational waves
\ba
\bald
\lb \partial_\tau^2 +2{\cal H}\partial_\tau +k^2  \rb h_{\lambda}(\bmk,\tau) &=  \frac{2}{m_p^2}~ \epsilon_{-\lambda}^{\alpha\beta}(\bmk) ~ \Pi^{\mu\nu}_{~~~\alpha\beta} (\bmk)~~~~~~~
\\ 
 &\times T_{\mu\nu} (\bmk, \tau) \,.
\label{h lambda eqn}
\eald
\ea
The solution of \eref{h lambda eqn} can be expressed as
\ba\label{h lambda}
\bald
h_\lambda(\bmk,\tau) &= h_\lambda^{\mbox{\scriptsize hom}}(\bmk,\tau) \\
&+ \frac{2}{m_p^2} \int d\tau' G_{k}(\tau,\tau') ~\epsilon_{-\lambda}^{\mu\nu}(\bmk) ~ T_{\mu\nu} (\bmk, \tau') \com~~~~
\eald
\ea
where $h_\lambda^{\mbox{\scriptsize hom}}(\bmk,\tau)$ is the homogeneous solution of \eref{h lambda eqn} and $ G_{k}(\tau,\tau')$  is the retarded Green's function of the differential operator on the left hand side of \eref{h lambda eqn}, i.e.,
\begin{eqnarray}
\left( \partial_\tau^2 +2{\cal H} \partial_\tau +k^2  \right)G_k(\tau,\tau')=\delta(\tau-\tau')\,.
\end{eqnarray}

We will be interested in the disparity between the left and right circular polarizations of the gravitational waves.  Since the homogeneous part contributes equally well to both the left and right handed modes, we will ignore this part in what follows.  Thus, the correlation function of the helical modes of the gravitational waves is given by
\ba
\bald
&\langle h_\lambda(\bmk,\tau) h_{\lambda'}(\bmk',\tau) \rangle =  \frac{4}{m_p^4} \int d\tau' d\tau'' \frac{1}{a(\tau')^2} \frac{1}{a(\tau'')^2} 
\\
&~~~~~~\times G_{k}(\tau,\tau') G_{k'}(\tau,\tau'') ~M_{\lambda \lambda'} (\bmk,\bmk';\tau,\tau',\tau'') \,,~~~~~~~~ 
\label{h correlator} 
\eald
\ea
and we defined
\ba\label{M}
\bald
M_{\lambda \lambda'} (\bmk,\bmk';\tau,\tau',\tau'') &\equiv  ~a(\tau')^2 a(\tau'')^2  \epsilon_{-\lambda}^{\mu\nu}(\bmk)\epsilon_{-\lambda'}^{\rho \sigma}(\bmk') ~~~~~~
\\
&\times\langle T_{\mu\nu} (\bmk, \tau')~  T_{\rho\sigma} (\bmk', \tau'') \rangle\,.~~~
\eald
\ea
The functions $M_{\lambda \lambda'}$ are simply the projections of the energy-momentum tensor correlators along the helicity eigenbasis. 
Thus, we can immediately see that an imbalance between $M_{++}$ and $M_{--}$ will cause a disparity between  $\langle h_+ h_+\rangle$ and $\langle h_- h_-\rangle$, and hence, the generation of chiral gravitational waves. The brackets $\langle\,\,\,\,\rangle$ in (\ref{h correlator}) can denote quantum, thermal, or stochastic averages. In the rest of this paper, we compute the expectation value of $M_{\lambda \lambda'}$ due to nonperturbative quantum production of chiral fermions in a time varying background, namely, in the FRW universe.

%===================================================
\subsection{Weyl Fermions in FRW Background}
%===================================================

The action for the free left-chiral massless Weyl fermions $\psi$ in a curved background reads
\begin{eqnarray}\label{main lagrangian}
S=\int d^4x \sqrt{-g}{\cal L}\,,\quad
{\cal L}=i\psi^\dagger e^{\mu}_a\bar \sigma^a D_\mu \psi\,,
\end{eqnarray}
where $e^{\mu}_a$ are the vielbeins, the latin letters denote the flat coordinates, and the Greek letters denote the curved coordinates. The covariant derivative is given by $D_\mu=\partial_\mu +\frac{1}{2}\sigma^{ab}\omega_{\mu ab}$, where $\omega_{\mu ab}$ are the spin connections, $\sigma^{ab}\equiv \frac{1}{4}\left[\sigma^a,\bar\sigma^b\right]$, $\sigma^a \equiv (1, \sigma^i)$, and $\bar \sigma^a \equiv (1, -\sigma^i)$,  where $\{\sigma^i\}$ are the Pauli matrices.
Using the conformal FRW metric to compute the spin connections and performing the change of variable $\psi=a^{3/2}\chi$ in (\ref{main lagrangian}) we obtain the Lagrangian
\begin{eqnarray}
{\cal L}=\frac{i}{a^4(\tau)}\chi^\dagger \delta_{a}^\mu\bar \sigma^a \partial_\mu\chi\,.
\end{eqnarray}
From now on it will be easier to stop distinguishing between the curved and flat coordinates, and we can just use the Greek letters to denote all quantities in flat Minkowski space, i.e., the above Lagrangian can be rewritten as 
${\cal L}=\frac{i}{a^4(\tau)}\chi^\dagger  \sigma^\mu \partial_\mu\chi$.

The Weyl fermions can be quantized using creation and annihilation operators as follows:
\ba
\bald
\chi(\bm x, \tau) = \inp ~ &\big[ u(\tau, \bmp) a(\bmp) e^{i \bmp \cdot \bmx}
\\
&+ v(\tau, \bmp) b^\dag(\bmp) e^{-i \bmp \cdot \bmx} \big] \xi_{-}( \bmp)\,, ~~~~~~
\label{chi L quantize 1}
\eald
\ea
where we use the shorthand notation $d^3 p/(2\pi)^3 \equiv \dbar^3 p$, and the functions $u$ and $v$ are the mode solutions of the Casimir of the representation $(\frac{1}{2},0)$ of the Poincare group. 
The creation and annihilation operators satisfy the anticommutation relations
\ba
\bald
\left\{ a(\bmp), a^\dag(\bmp') \right\} &=\left\{ b(\bmp), b^\dag(\bmp') \right\} = (2 \pi)^3 \delta^3 (\bmp - \bmp') 
\com \\
\left\{ a(\bmp), a(\bmp') \right\} &= \left\{ b(\bmp), b(\bmp') \right\} = \left\{ a(\bmp), b(\bmp') \right\} = 0 \,.~~~~~~
\eald
\ea
The spinor $\xi_{-}(\bmp)$ is the eigenstate of the helicity operator
\be
\bm \sigma \cdot \bmp ~\xi_{-}(\bmp) = - p~ \xi_{-}(\bmp) \com
\ee
where $p \equiv |\bmp|$, and it satisfies the relation
\ba
\label{xi relations}
\xi_{-}(\bmp) \xi_{-}^\dag(\bmp)= \frac{1}{2} \sigma^\mu \hat n_\mu (\bmp) \,,
\ea
where we have defined the unit 4-vector $\hat n^\mu$ as $\hat n^\mu (\bmp) \equiv (1, \bmp / (|\bmp|))$.  It is useful to factorize $e^{i \bmp \cdot \bmx}$ and rewrite Eq.~(\ref{chi L quantize})  as follows
\ba
\bald
\chi(\bm x, \tau) = \inp ~ &\big[ u(\tau, \bmp) a(\bmp) \xi_{-} (\bmp) 
\\
&+ v(\tau,-\bmp) b^\dag(-\bmp) \xi_{-}(-\bmp) \big]  e^{i \bmp \cdot \bmx} \per~~~~~~~
\label{chi L quantize}
\eald
\ea
It can be easily shown that the Weyl field operator $\chi$ annihilates a state with negative helicity particle and creates a state with positive helicity antiparticle, as required from the ${\cal C}{\cal P}{\cal T}$ theorem. 

The physical energy momentum tensor of the Weyl fermion is given by
\ba
\bald
&T_{\mu\nu} (\bmx,\tau) = \frac{1}{a(\tau)^2} \frac{i}{2} \big[ \chi^\dag(\bmx,\tau) \bar \sigma_{(\mu} \partial_{\nu) } \chi(\bmx,\tau)
\\
&~~~-\partial_{(\mu} \chi^\dag(\bmx,\tau)\bar \sigma_{\nu)} \chi(\bmx,\tau)\big]-\eta_{\mu\nu}\left[....\right]   \,,~~~~~
\label{TL}
\eald
\ea
where the term $\eta_{\mu\nu}\left[....\right]$ drops when we take the transverse part of the energy-momentum tensor, as we will do momentarily.  
In what follows we will need the Fourier transform of the energy momentum tensor to calculate the gravitational waves sourced by the Weyl fermions. Substituting Eq.~(\ref{chi L quantize}) into Eq.~(\ref{TL}), and then taking the Fourier transform, we obtain
\ba
\bald
T_{\mu\nu} (\bmk,\tau) &= \frac{1}{a(\tau)^2} \inx \inp \inp' ~L^\dag(\bmp',\tau)~~~~~~
\\
& \times \big[ \bar \sigma_{\mu} p_{\nu} + \bar \sigma_{\mu} p'_{\nu} \big]  
L(\bmp,\tau) ~e^{i (\bmp-\bmp'-\bmk) \cdot \bmx} \com
\label{TL Fourier}
\eald
\ea
and we defined the operator
\ba
\bald
L(\bmp,\tau) &\equiv \big[ u(\tau, \bmp) a(\bmp) \xi_{-} (\bmp) 
\\
&+ v(\tau, -\bmp) b^\dag(-\bmp) \xi_{-}(-\bmp) \big]\,.~~~~~
\label{L} 
\eald
\ea
Upon taking the integral over the 3-volume and  3-momentum $\bmp'$, we finally obtain the Fourier transform of the energy momentum tensor as
\ba
\bald
T_{\mu\nu} (\bmk,\tau) &= \frac{1}{a(\tau)^2} \inp ~L^\dag(\bmp-\bmk,\tau)
\\ 
&\times \lb \bar \sigma_{\mu} p_{\nu} + \bar \sigma_{\mu} (p-k)_{\nu} \rb  
L(\bmp,\tau)\,.
\label{TL k}
\eald
\ea

Having the energy-momentum tensor at hand, now we are ready to compute the two-point functions of the tensor perturbations. This is achieved in the next section.

%=============================================
\subsection{Chiral Gravitational Waves from Weyl Fermions}
%=============================================

The correlation function of the gravitational waves that result from  the production of left-chiral Weyl fermions can be found by substituting \eref{TL k}  into \eref{h correlator}. We will denote $M_{\lambda \lambda'} (\bmk,\bmk';\tau,\tau',\tau'')= M_{\lambda \lambda'}$ to reduce notational clutter. Then, the left handed Weyl fermions yield
\ba
\bald
&M_{\lambda \lambda'} =  \inp \inp' 
\langle L^\dag(\bmp-\bmk,\tau') \epsilon_{-\lambda}^{\mu\nu}(\bmk) \\
&\times \lb  \bar \sigma_{\mu} p_{\nu} + \bar \sigma_{\mu} (p-k)_{\nu} \rb  L(\bmp,\tau') ~ L^\dag(\bmp'-\bmk',\tau'')~~~~~~
\\
&\times \epsilon_{-\lambda'}^{\rho\sigma}(\bmk') \lb  \bar \sigma_{\rho} p'_{\sigma} + \bar \sigma_{\rho} (p'-k')_{\sigma} \rb  L(\bmp',\tau'') \rangle \label{ML} \per
\eald
\ea
The initial vacuum of the theory is defined as $a(\bmp)|0\rangle=b(\bmp)|0\rangle=0$. The time varying background will cause the production of fermions from vacuum,  and information about the background are encoded in the mode functions $u(\tau, \bmp)$ and $v(\tau, -\bmp)$. In this section we keep our formalism general enough and we do not specify a particular particle physics model that can lead to the production of left-chiral Weyl fermions.
Using the explicit forms of the $L(\bmp, \tau)$ operator given in \eref{L}, computing the vacuum expectation values of operators of the form 
\ba
\bald
&\bra{0} b(\bmk-\bmp) a(\bmp) a^\dag(\bmp'-\bmk') b^\dag(-\bmp') \ket{0} = (2\pi)^6 
\\
&~~~~~~~~~~~~~~~~~~~~~~~\times\delta^3(\bmk+\bmk') \delta^3(\bmp'-\bmp+\bmk)   \com~~~
\eald
\ea
ignoring  the contribution from the zero point fluctuations, and integrating over the 3-momentum $\bmp'$, we find after some algebra 
\ba
\bald
M_{\lambda \lambda'} &=  (2\pi)^3 \delta^3(\bmk+\bmk') \inp~ u(\tau', \bmp) u^*(\tau'', \bmp)~~~~~
\\
&\times v(\tau'',\bmk-\bmp) v^*(\tau',\bmk-\bmp) ~\beta_{\lambda \lambda'} (\bmk, \bmp)\,,
\eald
\ea
where we defined 
\ba
\bald
&\beta_{\lambda \lambda'} (\bmk, \bmp) \equiv
\lb \xi_{-}^\dag(\bmk - \bmp) ~\epsilon_{-\lambda}^{\mu}(\bmk) \bar \sigma_\mu  ~ \epsilon_{-\lambda}^{\nu}(\bmk) p_\nu ~ \xi_{-}(\bmp)  \rb ~~~~~~
\\
&~~~~~\times \lb \xi_{-}^\dag(\bmp) ~\epsilon_{-\lambda'}^{\rho}(-\bmk) \bar \sigma_\rho  ~ \epsilon_{-\lambda'}^{\sigma}(-\bmk) p_\sigma ~ \xi_{-}(\bmk - \bmp)  \rb\,.
\eald
\ea
Making use of  the completeness relation, \eref{xi relations}, we obtain
\ba
\bald
\beta_{\lambda \lambda'} (\bmk, \bmp) &= \frac{1}{4}~ \epsilon_{-\lambda} (\bmk) \cdot p ~~ \epsilon_{\lambda'} (\bmk) \cdot p ~\epsilon_{-\lambda}^{\mu}(\bmk) \epsilon_{\lambda'}^{\rho}(\bmk) ~~~~~~~
\\
&\times \hat n^\gamma (\bmp)~ \hat n^\alpha (\bmk-\bmp) ~ {\rm Tr}\lb \bar \sigma_\mu \sigma_\gamma \bar \sigma_\rho \sigma_\alpha \rb
\,.
\eald
\ea
Then, using the trace identity $ {\rm Tr}\lb \bar \sigma_\mu \sigma_\gamma \bar \sigma_\rho \sigma_\alpha \rb=2 \lb \eta_{\mu\gamma} \eta_{\rho\alpha} - \eta_{\mu\rho} \eta_{\gamma\alpha}+\eta_{\mu\alpha} \eta_{\gamma\rho} - i {\epsilon}_{\mu\gamma\rho\alpha} \rb$, where $\epsilon_{0123}=-1$, and fixing the coordinate system by choosing $\hat \bmp= (\sin\theta \cos\phi, \sin\theta\sin\phi, \cos\theta)$, we finally obtain the helical mode correlators that are sourced by the left-chiral Weyl fermions
\ba
\bald
&\langle h_\lambda(\bmk,\tau) h_{\lambda'}(\bmk',\tau) \rangle_\L =\delta_{\lambda\lambda'}~  \frac{\delta^3(\bmk+\bmk')}{m_p^4}  \int d^3 p~  p^2 ~d\tau'~ d\tau''
\\
&~~~ \times a(\tau')^{-2} a(\tau'')^{-2} G_{k}(\tau,\tau') G_{k}(\tau,\tau'')
\\
&~~~\times u(\tau', \bmp) u^*(\tau'', \bmp) v(\tau'',\bmk-\bmp) v^*(\tau',\bmk-\bmp)
\\
&~~~\times \sin^2\theta ~\Bigg[1 +  \frac{(\lambda+\lambda')}{2} \cos\theta
\\ 
&~~~~~~~~~~~~~~+\left(\cos\theta + \frac{(\lambda+\lambda')}{2} \right) \frac{p\cos\theta - k}{\sqrt{p^2 + k^2 - 2 p k \cos\theta}} \Bigg]
 \,. \label{h correlator final} \\
\eald
\ea
Similarly, one can also follow the same procedure to obtain an expression for the correlators due to contribution from right-handed Weyl fermions:
\ba
\bald
&\langle h_\lambda(\bmk,\tau) h_{\lambda'}(\bmk',\tau) \rangle_\R =\delta_{\lambda\lambda'}~  \frac{\delta^3(\bmk+\bmk')}{m_p^4}  \int d^3 p~  p^2 ~d\tau'~ d\tau''
\\
&~~~\times a(\tau')^{-2} a(\tau'')^{-2} G_{k}(\tau,\tau') G_{k}(\tau,\tau'')
\\
&~~~\times u(\tau', \bmp) u^*(\tau'', \bmp) v(\tau'',\bmk-\bmp) v^*(\tau',\bmk-\bmp)
\\
&~~~\times \sin^2\theta ~\Bigg[1 -  \frac{(\lambda+\lambda')}{2} \cos\theta 
\\
&~~~~~~~~~~~~~~+\left(\cos\theta - \frac{(\lambda+\lambda')}{2} \right) \frac{p\cos\theta - k}{\sqrt{p^2 + k^2 - 2 p k \cos\theta}} \Bigg] \per
\label{right h correlator} 
\eald
\ea
Equations (\ref{h correlator final}) and (\ref{right h correlator}) are the main results of our work. They reflect the fact that the production of fermions, from vacuum, with a definite helicity is accompanied by the generation of helical gravitational waves. However, we warn the reader that these equations are contaminated by ultraviolet (UV) divergences that have to be exorcised before making sense of their physical significance. Unlike the bosonic UV divergences, UV divergences of fermions in FRW background is a subtle topic that has been discussed in a few works, see, e.g., \cite{Landete:2013axa,Landete:2013lpa,delRio:2017iib}, and in the context of GW production in \cite{Enqvist:2012im,Figueroa:2013vif}. In this work we do not try to follow a rigorous procedure to regularize expressions (\ref{h correlator final}) and (\ref{right h correlator}). Instead, we follow a more phenomenological method, as we briefly discuss in Section  (\ref{sec:inflation}).

So far, our results did not depend on a specific particle physics model that can lead to the production of fermions with a definite helcity. In the next section, we examine the possibility of this scenario in a model of pseudoscalar inflation.

%=================================================================================
\section{Axion Coupling to Fermions and the Production of Chiral Gravitational Waves}
\label{sec:inflation}
%=================================================================================

In the previous section we showed that the production of fermions with a definite helicity will accompany the generation of helical gravitational waves. Fermion production in the early Universe via the parametric resonance has been the topic of many publications, see, e.g., \cite{Greene:1998nh,Giudice:1999fb,Greene:2000ew,Peloso:2000hy,Berges:2010zv}. However, in all these examples the produced fermions are vector-like, and hence, they source GW which are left-right symmetric.

In this section we discuss a specific model that is capable of producing either right- or left-handed fermions during inflation \cite{Adshead:2015kza}, and hence, the generation of helical tensor modes. This is achieved by coupling a Dirac fermion to an axion $\phi$. Axions could have played a very important role during the early Universe. In fact, they are perfect candidates to build radiatively stable models of inflation \cite{Freese:1990rb}, thanks to their continuous shift symmetry (which is broken to a discrete shift symmetry due to instanton effects).  This shift symmetry will guarantee that the fermions coupling to axions is governed by a dimension-$5$ operator \cite{Kim:2008hd}.  The action of the model is given in terms of two left-chiral Weyl fermions, $\psi_{1}$ and $\psi_2$ which together make up a Dirac fermion\footnote{ In particular, a four-spinor Dirac fermion can be written as $\Psi\equiv \left[\begin{array}{c} \psi_1 \\\psi_2^\dagger  \end{array}\right].$}, by \cite{Adshead:2015kza}
\ba
\bald
&S=\int d^4x\sqrt{-g}\Bigg[i\psi_1^\dagger e^{\mu}_a\bar \sigma^a D_\mu \psi_1+i\psi_2^\dagger e^{\mu}_a\bar \sigma^a D_\mu \psi_2
\\
&~~~ - m\left(\psi_1\psi_2+\psi_1^\dagger\psi_2^\dagger\right)  +\frac{C}{f}\partial_\mu \phi\left(\psi_1^\dagger e^{\mu}_a\bar \sigma^a D_\mu \psi_1\right)\Bigg] \com~~~~~~
\label{action Dirac fermion}
\eald
\ea
where $m$ is the fermion mass, $f$ is the axion constant, and $C$ is a coupling constant. This action is invariant under the U(1) global transformation $\psi_1\rightarrow e^{i\theta}\psi_1$ and $\psi_2\rightarrow e^{-i\theta}\psi_2$. Notice that the left-chiral Weyl fermions, $\psi_1$ and $\psi_2$, are eigenstates of the U(1) charge and that they have opposite charges under the U(1) symmetry. 

In order to study the generation of fermions in FRW background one defines $\chi_{1,2}\equiv \psi_{1,2}~a^{-3/2}$ and expands $\chi_1$ and $\chi_2$ in helicity eigensates as:
\ba
\bald
&\chi_1=\sum_{\lambda}\inp ~ \Bigg[ u(\tau, \bmp,\lambda)\xi_{\lambda}( \bmp) a(\bmp,\lambda) e^{i \bmp \cdot \bmx}
\\
&\qquad \qquad + v(\tau, \bmp,\lambda)\xi_{-\lambda}( \bmp) b^\dag(\bmp,\lambda) e^{-i \bmp \cdot \bmx} \Bigg] \,,
\\
&\chi_2=\sum_{\lambda}\inp ~ \Bigg[ u(\tau, \bmp,\lambda)\xi_{\lambda}( \bmp) b(\bmp,\lambda) e^{i \bmp \cdot \bmx}
\\
&\qquad \qquad  + v(\tau, \bmp,\lambda)\xi_{-\lambda}( \bmp) a^\dag(\bmp,\lambda) e^{-i \bmp \cdot \bmx} \Bigg] \per~~~~~~
\label{Dirac fermion expansion}
\eald
\ea
Notice that unlike the massless case, here both helicities, $\lambda=\pm$, are present in the expansion since a massive Weyl fermion does not have a definite helicity. Next, one substitutes the expansion (\ref{Dirac fermion expansion}) into the equations of motion, that result from varying the action (\ref{action Dirac fermion}) with respect to $\psi_{1,2}$, and studies the time evolution of the mode functions $u$ and $v$ assuming that one starts from the Bunch-Davies vacuum in the far past. This procedure was carried out and  thoroughly investigated in \cite{Adshead:2015kza}, during both the inflationary and preheating eras,  where the axion played the role of the inflaton. The conclusion is that during inflation and in the limit $H\gg m$ only one of the helicities, either left or right, of both fields, $\psi_1$ and $\psi_2$, will be generated depending on the sign of the parameter $\vartheta\equiv -\frac{C \dot \phi}{f H}$. During inflation we have to a very good approximation $\dot \phi=\mbox{constant}$ and hence the parameter $\vartheta$ stays almost constant. Taking $\vartheta<0$, $|\vartheta|\gg1$, and $m\ll H$, it was shown in \cite{Adshead:2015kza} that the particle production of helicity $\lambda=-$ is enhanced, while the production of helicity $\lambda=+$ is suppressed, i.e., $n_{\lambda=-}\cong 1$ and $n_{\lambda=+}\cong 0$, where $n$ is the fermion number density.\footnote{Notice that the nonpertutrbative production of fermions is blocked by the Pauli exclusion principle and hence we should expect $n\leq 1$.} Therefore, the physical picture is that the rolling inflaton (axion) leads to the production of fermions with negative helicity\footnote{Fermions of both positive and negative charges with respect to U(1) are equally produced.}, which in turn breaks the macroscopic parity of the space.  The mode functions that correspond to the helicity $\lambda=-$  are given by \cite{Adshead:2015kza}
\begin{eqnarray}
\nonumber
u(\tau,\bmp,-)&=&\frac{e^{\frac{\pi}{2}\vartheta}}{\sqrt{-2p\tau}}W_{\frac{1}{2}+i\vartheta,i\vartheta}\left(-2i p \tau\right)\,,\\
v(\tau,\bmp,+)&=&\frac{e^{-\frac{\pi}{2}\vartheta}}{\sqrt{-2p\tau}}W^*_{\frac{1}{2}-i\vartheta,i\vartheta}\left(-2i p \tau\right)\,,
\end{eqnarray}
where $W_{\kappa,\mu}$ are the Whittaker functions.
These are  the explicit forms of the mode functions $u$ and $v$ of the left-chiral Weyl fermions that appear in Eq.~(\ref{chi L quantize 1}). Both of the mode functions $u(\tau,\bmp,-)$ and $v(\tau,\bmp,+)$ accompany the negative helicity spinor $\xi_-$ as can be seen from \eref{Dirac fermion expansion}.

\begin{figure}[h] %  figure placement: here, top, bottom, or page
\centering
\includegraphics[width=.45\textwidth]{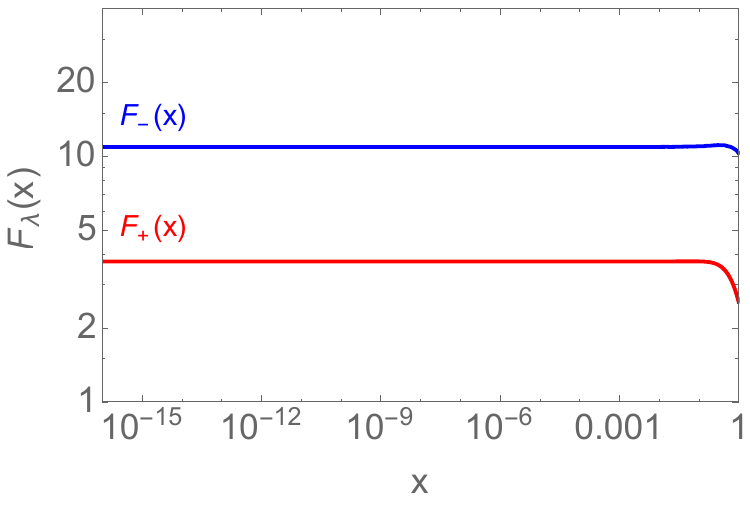} 
\caption{The numerical evaluation of the function ${\cal F}_{\lambda}(x)$ in \eref{F function} for both positive and negative helicities. Since fermions with negative helicity are produced from vaccum, we expect that the amplitude of $\langle h_-h_- \rangle$ to be bigger than that of $\langle h_+h_+ \rangle$, which is clear from the figure. We also find that the spectrum is almost scale invariant with a slight red tilt.  }
\label{fig:red spectrum}
\end{figure}

Now, we are in a position to calculate the correlators $\langle h_\lambda(\bmk,\tau) h_{\lambda'}(\bmk',\tau) \rangle$ from \eref{h correlator final}. The Green's function $G_k(\tau,\tau')$ in quasi de-Sitter background reads \cite{Sorbo:2011rz}
\ba
\bald
G_{k}(\tau,\tau') = \frac{1}{k^3 \tau'^2} \Bigg[ &(1+k^2 \tau \tau') \sin k(\tau-\tau') ~~~~~~~~~~
\\
&+ k(\tau'-\tau) \cos k(\tau-\tau') \Bigg]
\eald
\ea
for $\tau > \tau'$ and $G_{k}(\tau,\tau') =0$ for $\tau < \tau'$.
Making the change of variables $p=ky$, $k\tau'=x'$, and $k\tau''=x''$ in (\ref{h correlator final}), making use of $a=-\frac{1}{\tau H}$, and taking into account that we have two left-chiral Weyl fermions,  we find at the end of inflation, i.e. at $\tau\rightarrow -1/H$
\begin{eqnarray}
\langle h_\lambda(\bmk,\tau) h_{\lambda}(\bmk',\tau) \rangle=\frac{\delta^3(\bmk+\bmk')}{k^3}\frac{H^4}{m_p^4}{\cal F}_{\lambda}\left(\frac{k}{H}\right)\,,~~~~
\end{eqnarray}
where
\ba
\bald
&{\cal F}_{\lambda}\left( x\right) = \pi\int_0^\infty dy \int_{-\infty}^{-x}dx'\int_{-\infty}^{-x}dx''\int_0^\pi d\theta~~ x'x'' 
\\
&~~~\times{\cal G}(x,x'){\cal G}(x,x'') \frac{y^3 \sin^3\theta }{\sqrt{1+y^2-2y\cos\theta}}
\\
&~~~\times W_{\frac{1}{2}+i\vartheta, i\vartheta}\left(-2iyx'\right)W^*_{\frac{1}{2}+i\vartheta, i\vartheta}\left(-2iyx''\right)
\\
&~~~\times W_{\frac{1}{2}-i\vartheta, i\vartheta}\left(-2ix'\sqrt{1+y^2-2y\cos\theta}\right)
\\
&~~~\times W^*_{\frac{1}{2}-i\vartheta, i\vartheta}
\\
&~~~\times \left(-2ix''\sqrt{1+y^2-2y\cos\theta}\right) 
\\
&~~~ \times \left[1+\lambda\cos\theta+\frac{\left(\cos\theta+\lambda\right)\left(y\cos\theta-1\right)}{\sqrt{1+y^2-2y\cos\theta}}  \right]\,,
\label{F function}
\eald
\ea
and
\ba
\bald
{\cal G}(x,x') &=- \frac{1}{x'^2} \Big[ (1-xx') \sin (x+x') ~~~~~
\\
&- (x+x') \cos (x+x') \Big] \,.
\eald
\ea
In the limit $x \rightarrow -\infty$ the  Whittaker function behaves as $\mbox{lim}_{x\rightarrow \infty}W_{\frac{1}{2}+i\theta, i\theta}(ix)\approx (ix)^{\frac{1}{2}+i\theta}e^{-ix/2}$. Thus, the function ${\cal F}_{\lambda}$ receives an infinite contribution from the vacuum modes and it needs to be regularized. Here, we do not attempt to follow a rigorous regularization scheme\footnote{ See, e.g., \cite{Figueroa:2013vif} for a more sophisticated regularization scheme that can be used for fermion production in FRW background.}. Instead, we adopt a heuristic, yet a physical, method to cut off the divergent integrals. At the end of inflation, $a\approx 1$,  modes with wave numbers $p> H$ are deep in the UV, and they do not have enough time to get on shell. Hence, the integral over $p$ in (\ref{h correlator final}) should be cut off at momentum $ p\approx H$, i.e., the $y$ integral  should be cut off at values\footnote{In fact, the cutoff should depend on $\vartheta$ since the amplified modes have momenta of order $p\simeq \vartheta H$. Hence, the integral over $y$ in \eref{F function} should range from $0$ to $\vartheta/x \gtrsim1$. This dependence on $\vartheta$ is expected to appear in a more sophisticated regularization method, which is not captured by our scheme.} of $y \geq1$. Next, we observe that the  Whittaker function $W_{\frac{1}{2}+i\vartheta, i\vartheta}(ix)$ behaves as $\left(i x\right)^{\frac{1}{2}-i\theta}$ in the limit $x\rightarrow 0$.  Thus, the only difference between the two limits $x \rightarrow 0$ and $x \rightarrow \infty$ of the Whittaker function is the phase $e^{-i\frac{x}{2}}$, which causes the integrand to suffer from rapid oscillations. This oscillatory behavior is associated with the vacuum fluctuations deep in the UV. Therefore, we can set the phase to zero in our approximation of the Whittaker function in the limit $x\rightarrow \infty$, since anyway we are cutting off the integral in the UV. Thus, we can approximate the Whittaker function $W_{\frac{1}{2}+i\vartheta, i\vartheta}(ix)$ by $\left(i x\right)^{\frac{1}{2}-i\theta}$ throughout the entire evaluation of the integral. In addition, in order to efficiently cut off the integrals, we multiply every function $W_{\frac{1}{2}+i\vartheta, i\vartheta}(ix)$ by $e^{-\epsilon |x|}$, $\epsilon >0$. This exponential  factor will force the integrand to vanish when the argument of the Whittaker function becomes of order $1$ or bigger. After all, we know on physical grounds that most of the fermion production  occurs when $p |\tau| \rightarrow 0$, i.e., when the argument of the Whittaker function approaches zero. We used $\epsilon =0.5$ in our numerical integration shown in Fig.~\ref{fig:red spectrum}, and verified that the spectral shape is insensitive to the value of $\epsilon$. 

Making the series of the above mentioned approximations in (\ref{F function}), we find for the superhorizon modes, $k/H\ll1$, the power spectrum 
\ba
\bald
{\cal P}^-&\equiv\langle h_-(\bmk,\tau) h_{-}(\bmk,\tau) \rangle \cong\frac{H^2}{\pi^2 m_p^2}\left[1+108\frac{H^2}{m_p^2}    \right] \com ~~~~~~
\\
{\cal P}^+&\equiv\langle h_+(\bmk,\tau) h_{+}(\bmk,\tau) \rangle\cong \frac{H^2}{\pi^2 m_p^2}\left[1+37\frac{H^2}{m_p^2}    \right] \com
\eald
\ea
where we restored the homogenous part. Then,  the handedness of the tensor mode  is given by
\begin{eqnarray}
\Delta \chi=\frac{{\cal P}^--{\cal P}^+}{{\cal P}^- + {\cal P}^+}\cong 35\frac{H^2}{m_p^2} \per
\end{eqnarray}
Taking $H\leq 10^{13}$ GeV, we find that $\Delta \chi \leq 10^{-8}$, which is not in reach within the current CMB polarization experiments. Such a small signal is expected since the production of fermions is blocked by the Pauli exclusion principle.

We also evaluated the integral (\ref{F function}) for values of $k$ in the range $0<k/H<1$ to find that the spectrum is almost scale invariant with a slight red tilt as can be seen in Fig.~\ref{fig:red spectrum}. This scale invariance can be quantified by writing either ${\cal P}^+$ or ${\cal P}^-$ as
\begin{eqnarray}
\frac{H^2}{\pi^2 m_p^2}\left[1+C\frac{H^2}{m_p^2}\left(\frac{k}{k_p}\right)^{-n}\right]\,, 
\label{quantifying scale invariance}
\end{eqnarray} 
where $C$ is a constant, $k_p$ is a reference wave vector, and $n$ is an index that quantifies the deviation from scale invariance.  The value of $n$ can be extracted from our numerical calculations to find with a very good approximation $n\approx 10^{-4}$. Such nearly scale-invariant spectrum was also found in the chiral GW that accompany the nonperturbative production of helical photons due to their coupling to axions\footnote{The spectral index of the tensor modes  in \cite{Sorbo:2011rz} was not calculated. The spectral index of the scalar mode in the same model, however, was given in \cite{Anber:2009ua}, which indicated that the scalar spectrum is also nearly scalar invariant.}, \cite{Sorbo:2011rz}. Other models, however, predict a large deviation from scale invariance, see, e.g., \cite{Domcke:2016bkh}, and hence, the spectral index $n$ can be used to distinguish between various models\footnote{For instance see Refs.~\cite{Cook:2011hg,Barnaby:2011qe,Crowder:2012ik,Barnaby:2012xt,Anber:2012du,Maleknejad:2016qjz,Bartolo:2016ami} that study GW in various models.}.

%=================================================================================
\section{Summary and Discussion}
\label{sec:discussion}
%=================================================================================

In this work, we reported on a new mechanism to generate chiral gravitational waves from the imbalance between left and right-handed fermions. This imbalance breaks $\cal P$ and leads to an enhancement of a certain helicity mode of gravitational waves. In particular, we showed that the nonperturbative production of chiral Weyl fermions in a time varying background is accompanied by tensor perturbations of preferred helicity, see \eref{h correlator final} and \eref{right h correlator} for left and right-chiral fermions, respectively. This mechanism can be generalized to any process that creates an asymmetry between fermions of different helicities.

We also studied the generation of chiral GW from the production of fermions with a definite helicity in a model of pseudoscalar inflation \cite{Adshead:2015kza}. We calculated the power spectrum of the chiral components of the gravitational waves produced in this model. Chiral gravitational waves can be detected either indirectly using $\cal P$-odd TB and EB CMB correlators, which would otherwise be zero in the absence of $\cal P$ breaking, or using ground and space-based interferometers with polarization capabilities. In particular, we found that the amplitude of the difference between the chiral components of the superhorizon GW modes is not in reach within the current CMB polarization experiments \cite{Gerbino:2016mqb}. However, the situation  changes if a large number of chiral fermions  ${\cal N}$ is produced during inflation. Provided that the production of a large number of fermions does not backreact on inflation,  the handedness of the tensor mode is
\begin{eqnarray}
\Delta \chi\cong {\cal N}\frac{H^2}{m_p^2}\,.
\end{eqnarray}
Thus, a large number of fermions\footnote{Large number of fermions can be realized in models that invoke large number of sectors, for instance, see the recent discussion of the so called ${\cal N}$Naturalness \cite{Arkani-Hamed:2016rle}.}, ${\cal N} \approx 10^{8}$  for $H\approx 10^{13}$ GeV, is needed in order to result in a detectable signal in current CMB polarization experiments. Also, we found that the generated chiral GW are almost scale invariant with a slight red tilt. It remains to be studied whether subhorizon modes of these chiral GW can be directly detected in ground and space-based interferometers. 

Particle production during inflation and radiation dominated eras can leave features in the primordial GW spectrum. Recently, there has been an interest in the study of models that can lead to the generation of gravitational waves that might be detected in future ground and space-based interferometers \cite{Figueroa:2013vif,Figueroa:2016ojl,Cook:2011hg}. However, most of these studies have focused on the magnitude of the GW rather than their polarization. It will be interesting to study the possibility of detecting the polarization of the primordial GW in these experiments. Applications of the mechanism we reported on in this work to various processes in the early Universe as well as the possibility of detecting chiral GW in future experiments are under our current investigation and will appear elsewhere.

%=================================================================================
\acknowledgments

We would like to thank Peter Adshead and Lorenzo Sorbo for useful discussions. M.A. has been supported in part by the Swiss National Science Foundation and in part by M.J. Murdock Charitable Trust. E.S. is supported by the Swiss National Science Foundation and Alexander von Humboldt Foundation. 

%===================================================================================
\bibliographystyle{apsrev4-1}
\bibliography{chiralgravitons_refs}

\begin{thebibliography}{67}%
\makeatletter
\providecommand \@ifxundefined [1]{%
 \@ifx{#1\undefined}
}%
\providecommand \@ifnum [1]{%
 \ifnum #1\expandafter \@firstoftwo
 \else \expandafter \@secondoftwo
 \fi
}%
\providecommand \@ifx [1]{%
 \ifx #1\expandafter \@firstoftwo
 \else \expandafter \@secondoftwo
 \fi
}%
\providecommand \natexlab [1]{#1}%
\providecommand \enquote  [1]{``#1''}%
\providecommand \bibnamefont  [1]{#1}%
\providecommand \bibfnamefont [1]{#1}%
\providecommand \citenamefont [1]{#1}%
\providecommand \href@noop [0]{\@secondoftwo}%
\providecommand \href [0]{\begingroup \@sanitize@url \@href}%
\providecommand \@href[1]{\@@startlink{#1}\@@href}%
\providecommand \@@href[1]{\endgroup#1\@@endlink}%
\providecommand \@sanitize@url [0]{\catcode `\\12\catcode `\$12\catcode
  `\&12\catcode `\#12\catcode `\^12\catcode `\_12\catcode `\%12\relax}%
\providecommand \@@startlink[1]{}%
\providecommand \@@endlink[0]{}%
\providecommand \url  [0]{\begingroup\@sanitize@url \@url }%
\providecommand \@url [1]{\endgroup\@href {#1}{\urlprefix }}%
\providecommand \urlprefix  [0]{URL }%
\providecommand \Eprint [0]{\href }%
\providecommand \doibase [0]{http://dx.doi.org/}%
\providecommand \selectlanguage [0]{\@gobble}%
\providecommand \bibinfo  [0]{\@secondoftwo}%
\providecommand \bibfield  [0]{\@secondoftwo}%
\providecommand \translation [1]{[#1]}%
\providecommand \BibitemOpen [0]{}%
\providecommand \bibitemStop [0]{}%
\providecommand \bibitemNoStop [0]{.\EOS\space}%
\providecommand \EOS [0]{\spacefactor3000\relax}%
\providecommand \BibitemShut  [1]{\csname bibitem#1\endcsname}%
\let\auto@bib@innerbib\@empty
%</preamble>
\bibitem [{\citenamefont {Ade}\ \emph {et~al.}(2016)\citenamefont {Ade} \emph
  {et~al.}}]{Array:2015xqh}%
  \BibitemOpen
  \bibfield  {author} {\bibinfo {author} {\bibfnamefont {P.~A.~R.}\
  \bibnamefont {Ade}} \emph {et~al.} (\bibinfo {collaboration} {BICEP2, Keck
  Array}),\ }\href {\doibase 10.1103/PhysRevLett.116.031302} {\bibfield
  {journal} {\bibinfo  {journal} {Phys. Rev. Lett.}\ }\textbf {\bibinfo
  {volume} {116}},\ \bibinfo {pages} {031302} (\bibinfo {year} {2016})},\
  \Eprint {http://arxiv.org/abs/1510.09217} {arXiv:1510.09217 [astro-ph.CO]}
  \BibitemShut {NoStop}%
%%CITATION = ARXIV:1510.09217;%%
\bibitem [{\citenamefont {Ahmed}\ \emph {et~al.}(2014)\citenamefont {Ahmed}
  \emph {et~al.}}]{Ahmed:2014ixy}%
  \BibitemOpen
  \bibfield  {author} {\bibinfo {author} {\bibfnamefont {Z.}~\bibnamefont
  {Ahmed}} \emph {et~al.} (\bibinfo {collaboration} {BICEP3}),\ }\bibfield
  {booktitle} {\emph {\bibinfo {booktitle} {{Proceedings, SPIE Astronomical
  Telescopes + Instrumentation 2014: Millimeter, Submillimeter, and
  Far-Infrared Detectors and Instrumentation for Astronomy VII}}},\ }\href
  {\doibase 10.1117/12.2057224} {\bibfield  {journal} {\bibinfo  {journal}
  {Proc. SPIE Int. Soc. Opt. Eng.}\ }\textbf {\bibinfo {volume} {9153}},\
  \bibinfo {pages} {91531N} (\bibinfo {year} {2014})},\ \Eprint
  {http://arxiv.org/abs/1407.5928} {arXiv:1407.5928 [astro-ph.IM]} \BibitemShut
  {NoStop}%
%%CITATION = ARXIV:1407.5928;%%
\bibitem [{\citenamefont {Ade}\ \emph {et~al.}(2014)\citenamefont {Ade} \emph
  {et~al.}}]{Ade:2014afa}%
  \BibitemOpen
  \bibfield  {author} {\bibinfo {author} {\bibfnamefont {P.~A.~R.}\
  \bibnamefont {Ade}} \emph {et~al.} (\bibinfo {collaboration} {POLARBEAR}),\
  }\href {\doibase 10.1088/0004-637X/794/2/171} {\bibfield  {journal} {\bibinfo
   {journal} {Astrophys. J.}\ }\textbf {\bibinfo {volume} {794}},\ \bibinfo
  {pages} {171} (\bibinfo {year} {2014})},\ \Eprint
  {http://arxiv.org/abs/1403.2369} {arXiv:1403.2369 [astro-ph.CO]} \BibitemShut
  {NoStop}%
%%CITATION = ARXIV:1403.2369;%%
\bibitem [{\citenamefont {Keisler}\ \emph {et~al.}(2015)\citenamefont {Keisler}
  \emph {et~al.}}]{Keisler:2015hfa}%
  \BibitemOpen
  \bibfield  {author} {\bibinfo {author} {\bibfnamefont {R.}~\bibnamefont
  {Keisler}} \emph {et~al.} (\bibinfo {collaboration} {SPT}),\ }\href {\doibase
  10.1088/0004-637X/807/2/151} {\bibfield  {journal} {\bibinfo  {journal}
  {Astrophys. J.}\ }\textbf {\bibinfo {volume} {807}},\ \bibinfo {pages} {151}
  (\bibinfo {year} {2015})},\ \Eprint {http://arxiv.org/abs/1503.02315}
  {arXiv:1503.02315 [astro-ph.CO]} \BibitemShut {NoStop}%
%%CITATION = ARXIV:1503.02315;%%
\bibitem [{\citenamefont {Niemack}\ \emph {et~al.}(2010)\citenamefont {Niemack}
  \emph {et~al.}}]{Niemack:2010wz}%
  \BibitemOpen
  \bibfield  {author} {\bibinfo {author} {\bibfnamefont {M.~D.}\ \bibnamefont
  {Niemack}} \emph {et~al.},\ }\href {\doibase 10.1117/12.857464} {\bibfield
  {journal} {\bibinfo  {journal} {Proc. SPIE Int. Soc. Opt. Eng.}\ }\textbf
  {\bibinfo {volume} {7741}},\ \bibinfo {pages} {77411S} (\bibinfo {year}
  {2010})},\ \Eprint {http://arxiv.org/abs/1006.5049} {arXiv:1006.5049
  [astro-ph.IM]} \BibitemShut {NoStop}%
%%CITATION = ARXIV:1006.5049;%%
\bibitem [{\citenamefont {Abbott}\ \emph
  {et~al.}(2016{\natexlab{a}})\citenamefont {Abbott} \emph
  {et~al.}}]{Abbott:2016blz}%
  \BibitemOpen
  \bibfield  {author} {\bibinfo {author} {\bibfnamefont {B.~P.}\ \bibnamefont
  {Abbott}} \emph {et~al.} (\bibinfo {collaboration} {Virgo, LIGO
  Scientific}),\ }\href {\doibase 10.1103/PhysRevLett.116.061102} {\bibfield
  {journal} {\bibinfo  {journal} {Phys. Rev. Lett.}\ }\textbf {\bibinfo
  {volume} {116}},\ \bibinfo {pages} {061102} (\bibinfo {year}
  {2016}{\natexlab{a}})},\ \Eprint {http://arxiv.org/abs/1602.03837}
  {arXiv:1602.03837 [gr-qc]} \BibitemShut {NoStop}%
%%CITATION = ARXIV:1602.03837;%%
\bibitem [{\citenamefont {Abbott}\ \emph
  {et~al.}(2016{\natexlab{b}})\citenamefont {Abbott} \emph
  {et~al.}}]{Abbott:2016nmj}%
  \BibitemOpen
  \bibfield  {author} {\bibinfo {author} {\bibfnamefont {B.}~\bibnamefont
  {Abbott}} \emph {et~al.} (\bibinfo {collaboration} {Virgo, LIGO
  Scientific}),\ }\href {\doibase 10.1103/PhysRevLett.116.241103} {\bibfield
  {journal} {\bibinfo  {journal} {Phys. Rev. Lett.}\ }\textbf {\bibinfo
  {volume} {116}},\ \bibinfo {pages} {241103} (\bibinfo {year}
  {2016}{\natexlab{b}})},\ \Eprint {http://arxiv.org/abs/1606.04855}
  {arXiv:1606.04855 [gr-qc]} \BibitemShut {NoStop}%
%%CITATION = ARXIV:1606.04855;%%
\bibitem [{\citenamefont {Joyce}\ and\ \citenamefont
  {Shaposhnikov}(1997)}]{Joyce:1997uy}%
  \BibitemOpen
  \bibfield  {author} {\bibinfo {author} {\bibfnamefont {M.}~\bibnamefont
  {Joyce}}\ and\ \bibinfo {author} {\bibfnamefont {M.~E.}\ \bibnamefont
  {Shaposhnikov}},\ }\href {\doibase 10.1103/PhysRevLett.79.1193} {\bibfield
  {journal} {\bibinfo  {journal} {Phys. Rev. Lett.}\ }\textbf {\bibinfo
  {volume} {79}},\ \bibinfo {pages} {1193} (\bibinfo {year} {1997})},\ \Eprint
  {http://arxiv.org/abs/astro-ph/9703005} {arXiv:astro-ph/9703005 [astro-ph]}
  \BibitemShut {NoStop}%
%%CITATION = ASTRO-PH/9703005;%%
\bibitem [{\citenamefont {Brustein}\ and\ \citenamefont
  {Oaknin}(1999)}]{Brustein:1998du}%
  \BibitemOpen
  \bibfield  {author} {\bibinfo {author} {\bibfnamefont {R.}~\bibnamefont
  {Brustein}}\ and\ \bibinfo {author} {\bibfnamefont {D.~H.}\ \bibnamefont
  {Oaknin}},\ }\href {\doibase 10.1103/PhysRevLett.82.2628} {\bibfield
  {journal} {\bibinfo  {journal} {Phys. Rev. Lett.}\ }\textbf {\bibinfo
  {volume} {82}},\ \bibinfo {pages} {2628} (\bibinfo {year} {1999})},\ \Eprint
  {http://arxiv.org/abs/hep-ph/9809365} {arXiv:hep-ph/9809365 [hep-ph]}
  \BibitemShut {NoStop}%
%%CITATION = HEP-PH/9809365;%%
\bibitem [{\citenamefont {Giovannini}\ and\ \citenamefont
  {Shaposhnikov}(1998)}]{Giovannini:1997eg}%
  \BibitemOpen
  \bibfield  {author} {\bibinfo {author} {\bibfnamefont {M.}~\bibnamefont
  {Giovannini}}\ and\ \bibinfo {author} {\bibfnamefont {M.}~\bibnamefont
  {Shaposhnikov}},\ }\href {\doibase 10.1103/PhysRevD.57.2186} {\bibfield
  {journal} {\bibinfo  {journal} {Phys.Rev.}\ }\textbf {\bibinfo {volume}
  {D57}},\ \bibinfo {pages} {2186} (\bibinfo {year} {1998})},\ \Eprint
  {http://arxiv.org/abs/hep-ph/9710234} {arXiv:hep-ph/9710234 [hep-ph]}
  \BibitemShut {NoStop}%
%%CITATION = HEP-PH/9710234;%%
\bibitem [{\citenamefont {Giovannini}(2000{\natexlab{a}})}]{Giovannini:1999wv}%
  \BibitemOpen
  \bibfield  {author} {\bibinfo {author} {\bibfnamefont {M.}~\bibnamefont
  {Giovannini}},\ }\href {\doibase 10.1103/PhysRevD.61.063004} {\bibfield
  {journal} {\bibinfo  {journal} {Phys. Rev.}\ }\textbf {\bibinfo {volume}
  {D61}},\ \bibinfo {pages} {063004} (\bibinfo {year} {2000}{\natexlab{a}})},\
  \Eprint {http://arxiv.org/abs/hep-ph/9905358} {arXiv:hep-ph/9905358 [hep-ph]}
  \BibitemShut {NoStop}%
%%CITATION = HEP-PH/9905358;%%
\bibitem [{\citenamefont {Giovannini}(2000{\natexlab{b}})}]{Giovannini:1999by}%
  \BibitemOpen
  \bibfield  {author} {\bibinfo {author} {\bibfnamefont {M.}~\bibnamefont
  {Giovannini}},\ }\href {\doibase 10.1103/PhysRevD.61.063502} {\bibfield
  {journal} {\bibinfo  {journal} {Phys. Rev.}\ }\textbf {\bibinfo {volume}
  {D61}},\ \bibinfo {pages} {063502} (\bibinfo {year} {2000}{\natexlab{b}})},\
  \Eprint {http://arxiv.org/abs/hep-ph/9906241} {arXiv:hep-ph/9906241 [hep-ph]}
  \BibitemShut {NoStop}%
%%CITATION = HEP-PH/9906241;%%
\bibitem [{\citenamefont {Bamba}(2006)}]{Bamba:2006km}%
  \BibitemOpen
  \bibfield  {author} {\bibinfo {author} {\bibfnamefont {K.}~\bibnamefont
  {Bamba}},\ }\href {\doibase 10.1103/PhysRevD.74.123504} {\bibfield  {journal}
  {\bibinfo  {journal} {Phys. Rev.}\ }\textbf {\bibinfo {volume} {D74}},\
  \bibinfo {pages} {123504} (\bibinfo {year} {2006})},\ \Eprint
  {http://arxiv.org/abs/hep-ph/0611152} {arXiv:hep-ph/0611152 [hep-ph]}
  \BibitemShut {NoStop}%
%%CITATION = HEP-PH/0611152;%%
\bibitem [{\citenamefont {Bamba}\ \emph {et~al.}(2008)\citenamefont {Bamba},
  \citenamefont {Geng},\ and\ \citenamefont {Ho}}]{Bamba:2007hf}%
  \BibitemOpen
  \bibfield  {author} {\bibinfo {author} {\bibfnamefont {K.}~\bibnamefont
  {Bamba}}, \bibinfo {author} {\bibfnamefont {C.~Q.}\ \bibnamefont {Geng}}, \
  and\ \bibinfo {author} {\bibfnamefont {S.~H.}\ \bibnamefont {Ho}},\ }\href
  {\doibase 10.1016/j.physletb.2008.05.027} {\bibfield  {journal} {\bibinfo
  {journal} {Phys. Lett.}\ }\textbf {\bibinfo {volume} {B664}},\ \bibinfo
  {pages} {154} (\bibinfo {year} {2008})},\ \Eprint
  {http://arxiv.org/abs/0712.1523} {arXiv:0712.1523 [hep-ph]} \BibitemShut
  {NoStop}%
%%CITATION = ARXIV:0712.1523;%%
\bibitem [{\citenamefont {Fujita}\ and\ \citenamefont
  {Kamada}(2016)}]{Fujita:2016igl}%
  \BibitemOpen
  \bibfield  {author} {\bibinfo {author} {\bibfnamefont {T.}~\bibnamefont
  {Fujita}}\ and\ \bibinfo {author} {\bibfnamefont {K.}~\bibnamefont
  {Kamada}},\ }\href {\doibase 10.1103/PhysRevD.93.083520} {\bibfield
  {journal} {\bibinfo  {journal} {Phys. Rev.}\ }\textbf {\bibinfo {volume}
  {D93}},\ \bibinfo {pages} {083520} (\bibinfo {year} {2016})},\ \Eprint
  {http://arxiv.org/abs/1602.02109} {arXiv:1602.02109 [hep-ph]} \BibitemShut
  {NoStop}%
%%CITATION = ARXIV:1602.02109;%%
\bibitem [{\citenamefont {Domcke}\ \emph {et~al.}(2016)\citenamefont {Domcke},
  \citenamefont {Pieroni},\ and\ \citenamefont {Binétruy}}]{Domcke:2016bkh}%
  \BibitemOpen
  \bibfield  {author} {\bibinfo {author} {\bibfnamefont {V.}~\bibnamefont
  {Domcke}}, \bibinfo {author} {\bibfnamefont {M.}~\bibnamefont {Pieroni}}, \
  and\ \bibinfo {author} {\bibfnamefont {P.}~\bibnamefont {Binétruy}},\ }\href
  {\doibase 10.1088/1475-7516/2016/06/031} {\bibfield  {journal} {\bibinfo
  {journal} {JCAP}\ }\textbf {\bibinfo {volume} {1606}},\ \bibinfo {pages}
  {031} (\bibinfo {year} {2016})},\ \Eprint {http://arxiv.org/abs/1603.01287}
  {arXiv:1603.01287 [astro-ph.CO]} \BibitemShut {NoStop}%
%%CITATION = ARXIV:1603.01287;%%
\bibitem [{\citenamefont {Ben-Dayan}(2016)}]{Ben-Dayan:2016iks}%
  \BibitemOpen
  \bibfield  {author} {\bibinfo {author} {\bibfnamefont {I.}~\bibnamefont
  {Ben-Dayan}},\ }\href@noop {} {\  (\bibinfo {year} {2016})},\ \Eprint
  {http://arxiv.org/abs/1604.07899} {arXiv:1604.07899 [astro-ph.CO]}
  \BibitemShut {NoStop}%
%%CITATION = ARXIV:1604.07899;%%
\bibitem [{\citenamefont {Adshead}\ \emph {et~al.}(2016)\citenamefont
  {Adshead}, \citenamefont {Giblin}, \citenamefont {Scully},\ and\
  \citenamefont {Sfakianakis}}]{Adshead:2016iae}%
  \BibitemOpen
  \bibfield  {author} {\bibinfo {author} {\bibfnamefont {P.}~\bibnamefont
  {Adshead}}, \bibinfo {author} {\bibfnamefont {J.~T.}\ \bibnamefont {Giblin}},
  \bibinfo {author} {\bibfnamefont {T.~R.}\ \bibnamefont {Scully}}, \ and\
  \bibinfo {author} {\bibfnamefont {E.~I.}\ \bibnamefont {Sfakianakis}},\
  }\href@noop {} {\  (\bibinfo {year} {2016})},\ \Eprint
  {http://arxiv.org/abs/1606.08474} {arXiv:1606.08474 [astro-ph.CO]}
  \BibitemShut {NoStop}%
%%CITATION = ARXIV:1606.08474;%%
\bibitem [{\citenamefont {Kamada}\ and\ \citenamefont
  {Long}(2016)}]{Kamada:2016eeb}%
  \BibitemOpen
  \bibfield  {author} {\bibinfo {author} {\bibfnamefont {K.}~\bibnamefont
  {Kamada}}\ and\ \bibinfo {author} {\bibfnamefont {A.~J.}\ \bibnamefont
  {Long}},\ }\href@noop {} {\  (\bibinfo {year} {2016})},\ \Eprint
  {http://arxiv.org/abs/1606.08891} {arXiv:1606.08891 [astro-ph.CO]}
  \BibitemShut {NoStop}%
%%CITATION = ARXIV:1606.08891;%%
\bibitem [{\citenamefont {Anber}\ and\ \citenamefont
  {Sabancilar}(2015)}]{Anber:2015yca}%
  \BibitemOpen
  \bibfield  {author} {\bibinfo {author} {\bibfnamefont {M.~M.}\ \bibnamefont
  {Anber}}\ and\ \bibinfo {author} {\bibfnamefont {E.}~\bibnamefont
  {Sabancilar}},\ }\href {\doibase 10.1103/PhysRevD.92.101501} {\bibfield
  {journal} {\bibinfo  {journal} {Phys. Rev.}\ }\textbf {\bibinfo {volume}
  {D92}},\ \bibinfo {pages} {101501} (\bibinfo {year} {2015})},\ \Eprint
  {http://arxiv.org/abs/1507.00744} {arXiv:1507.00744 [hep-th]} \BibitemShut
  {NoStop}%
%%CITATION = ARXIV:1507.00744;%%
\bibitem [{\citenamefont {Alexander}\ \emph {et~al.}(2016)\citenamefont
  {Alexander}, \citenamefont {Cormack},\ and\ \citenamefont
  {Sims}}]{Alexander:2016moy}%
  \BibitemOpen
  \bibfield  {author} {\bibinfo {author} {\bibfnamefont {S.}~\bibnamefont
  {Alexander}}, \bibinfo {author} {\bibfnamefont {S.}~\bibnamefont {Cormack}},
  \ and\ \bibinfo {author} {\bibfnamefont {R.}~\bibnamefont {Sims}},\
  }\href@noop {} {\  (\bibinfo {year} {2016})},\ \Eprint
  {http://arxiv.org/abs/1606.05357} {arXiv:1606.05357 [astro-ph.CO]}
  \BibitemShut {NoStop}%
%%CITATION = ARXIV:1606.05357;%%
\bibitem [{\citenamefont {Alexander}\ \emph {et~al.}(2006)\citenamefont
  {Alexander}, \citenamefont {Peskin},\ and\ \citenamefont
  {Sheikh-Jabbari}}]{Alexander:2004us}%
  \BibitemOpen
  \bibfield  {author} {\bibinfo {author} {\bibfnamefont {S.~H.-S.}\
  \bibnamefont {Alexander}}, \bibinfo {author} {\bibfnamefont {M.~E.}\
  \bibnamefont {Peskin}}, \ and\ \bibinfo {author} {\bibfnamefont {M.~M.}\
  \bibnamefont {Sheikh-Jabbari}},\ }\href {\doibase
  10.1103/PhysRevLett.96.081301} {\bibfield  {journal} {\bibinfo  {journal}
  {Phys. Rev. Lett.}\ }\textbf {\bibinfo {volume} {96}},\ \bibinfo {pages}
  {081301} (\bibinfo {year} {2006})},\ \Eprint
  {http://arxiv.org/abs/hep-th/0403069} {arXiv:hep-th/0403069 [hep-th]}
  \BibitemShut {NoStop}%
%%CITATION = HEP-TH/0403069;%%
\bibitem [{\citenamefont {Adshead}\ and\ \citenamefont
  {Sfakianakis}(2016)}]{Adshead:2015jza}%
  \BibitemOpen
  \bibfield  {author} {\bibinfo {author} {\bibfnamefont {P.}~\bibnamefont
  {Adshead}}\ and\ \bibinfo {author} {\bibfnamefont {E.~I.}\ \bibnamefont
  {Sfakianakis}},\ }\href {\doibase 10.1103/PhysRevLett.116.091301} {\bibfield
  {journal} {\bibinfo  {journal} {Phys. Rev. Lett.}\ }\textbf {\bibinfo
  {volume} {116}},\ \bibinfo {pages} {091301} (\bibinfo {year} {2016})},\
  \Eprint {http://arxiv.org/abs/1508.00881} {arXiv:1508.00881 [hep-ph]}
  \BibitemShut {NoStop}%
%%CITATION = ARXIV:1508.00881;%%
\bibitem [{\citenamefont {Long}\ \emph {et~al.}(2014)\citenamefont {Long},
  \citenamefont {Sabancilar},\ and\ \citenamefont {Vachaspati}}]{Long:2013tha}%
  \BibitemOpen
  \bibfield  {author} {\bibinfo {author} {\bibfnamefont {A.~J.}\ \bibnamefont
  {Long}}, \bibinfo {author} {\bibfnamefont {E.}~\bibnamefont {Sabancilar}}, \
  and\ \bibinfo {author} {\bibfnamefont {T.}~\bibnamefont {Vachaspati}},\
  }\href {\doibase 10.1088/1475-7516/2014/02/036} {\bibfield  {journal}
  {\bibinfo  {journal} {JCAP}\ }\textbf {\bibinfo {volume} {1402}},\ \bibinfo
  {pages} {036} (\bibinfo {year} {2014})},\ \Eprint
  {http://arxiv.org/abs/1309.2315} {arXiv:1309.2315 [astro-ph.CO]} \BibitemShut
  {NoStop}%
%%CITATION = ARXIV:1309.2315;%%
\bibitem [{\citenamefont {Sabancilar}(2013)}]{Sabancilar:2013raa}%
  \BibitemOpen
  \bibfield  {author} {\bibinfo {author} {\bibfnamefont {E.}~\bibnamefont
  {Sabancilar}},\ }\href@noop {} {\  (\bibinfo {year} {2013})},\ \Eprint
  {http://arxiv.org/abs/1310.8632} {arXiv:1310.8632 [hep-th]} \BibitemShut
  {NoStop}%
%%CITATION = ARXIV:1310.8632;%%
\bibitem [{\citenamefont {Gluscevic}\ and\ \citenamefont
  {Kamionkowski}(2010)}]{Gluscevic:2010vv}%
  \BibitemOpen
  \bibfield  {author} {\bibinfo {author} {\bibfnamefont {V.}~\bibnamefont
  {Gluscevic}}\ and\ \bibinfo {author} {\bibfnamefont {M.}~\bibnamefont
  {Kamionkowski}},\ }\href {\doibase 10.1103/PhysRevD.81.123529} {\bibfield
  {journal} {\bibinfo  {journal} {Phys. Rev.}\ }\textbf {\bibinfo {volume}
  {D81}},\ \bibinfo {pages} {123529} (\bibinfo {year} {2010})},\ \Eprint
  {http://arxiv.org/abs/1002.1308} {arXiv:1002.1308 [astro-ph.CO]} \BibitemShut
  {NoStop}%
%%CITATION = ARXIV:1002.1308;%%
\bibitem [{\citenamefont {Hayama}\ \emph {et~al.}(2016)\citenamefont {Hayama},
  \citenamefont {Kuroda}, \citenamefont {Nakamura},\ and\ \citenamefont
  {Yamada}}]{Hayama:2016kmv}%
  \BibitemOpen
  \bibfield  {author} {\bibinfo {author} {\bibfnamefont {K.}~\bibnamefont
  {Hayama}}, \bibinfo {author} {\bibfnamefont {T.}~\bibnamefont {Kuroda}},
  \bibinfo {author} {\bibfnamefont {K.}~\bibnamefont {Nakamura}}, \ and\
  \bibinfo {author} {\bibfnamefont {S.}~\bibnamefont {Yamada}},\ }\href
  {\doibase 10.1103/PhysRevLett.116.151102} {\bibfield  {journal} {\bibinfo
  {journal} {Phys. Rev. Lett.}\ }\textbf {\bibinfo {volume} {116}},\ \bibinfo
  {pages} {151102} (\bibinfo {year} {2016})},\ \Eprint
  {http://arxiv.org/abs/1606.01520} {arXiv:1606.01520 [astro-ph.HE]}
  \BibitemShut {NoStop}%
%%CITATION = ARXIV:1606.01520;%%
\bibitem [{\citenamefont {Aasi}\ \emph {et~al.}(2014)\citenamefont {Aasi} \emph
  {et~al.}}]{Aasi:2014erp}%
  \BibitemOpen
  \bibfield  {author} {\bibinfo {author} {\bibfnamefont {J.}~\bibnamefont
  {Aasi}} \emph {et~al.} (\bibinfo {collaboration} {VIRGO, LIGO Scientific}),\
  }\href {\doibase 10.1103/PhysRevD.90.062010} {\bibfield  {journal} {\bibinfo
  {journal} {Phys. Rev.}\ }\textbf {\bibinfo {volume} {D90}},\ \bibinfo {pages}
  {062010} (\bibinfo {year} {2014})},\ \Eprint {http://arxiv.org/abs/1405.7904}
  {arXiv:1405.7904 [gr-qc]} \BibitemShut {NoStop}%
%%CITATION = ARXIV:1405.7904;%%
\bibitem [{\citenamefont {Cook}\ and\ \citenamefont
  {Sorbo}(2012)}]{Cook:2011hg}%
  \BibitemOpen
  \bibfield  {author} {\bibinfo {author} {\bibfnamefont {J.~L.}\ \bibnamefont
  {Cook}}\ and\ \bibinfo {author} {\bibfnamefont {L.}~\bibnamefont {Sorbo}},\
  }\href {\doibase 10.1103/PhysRevD.86.069901, 10.1103/PhysRevD.85.023534}
  {\bibfield  {journal} {\bibinfo  {journal} {Phys. Rev.}\ }\textbf {\bibinfo
  {volume} {D85}},\ \bibinfo {pages} {023534} (\bibinfo {year} {2012})},\
  \bibinfo {note} {[Erratum: Phys. Rev.D86,069901(2012)]},\ \Eprint
  {http://arxiv.org/abs/1109.0022} {arXiv:1109.0022 [astro-ph.CO]} \BibitemShut
  {NoStop}%
%%CITATION = ARXIV:1109.0022;%%
\bibitem [{\citenamefont {Choi}\ \emph {et~al.}(2000)\citenamefont {Choi},
  \citenamefont {Hwang},\ and\ \citenamefont {Hwang}}]{Choi:1999zy}%
  \BibitemOpen
  \bibfield  {author} {\bibinfo {author} {\bibfnamefont {K.}~\bibnamefont
  {Choi}}, \bibinfo {author} {\bibfnamefont {J.-c.}\ \bibnamefont {Hwang}}, \
  and\ \bibinfo {author} {\bibfnamefont {K.~W.}\ \bibnamefont {Hwang}},\ }\href
  {\doibase 10.1103/PhysRevD.61.084026} {\bibfield  {journal} {\bibinfo
  {journal} {Phys. Rev.}\ }\textbf {\bibinfo {volume} {D61}},\ \bibinfo {pages}
  {084026} (\bibinfo {year} {2000})},\ \Eprint
  {http://arxiv.org/abs/hep-ph/9907244} {arXiv:hep-ph/9907244 [hep-ph]}
  \BibitemShut {NoStop}%
%%CITATION = HEP-PH/9907244;%%
\bibitem [{\citenamefont {Lue}\ \emph {et~al.}(1999)\citenamefont {Lue},
  \citenamefont {Wang},\ and\ \citenamefont {Kamionkowski}}]{Lue:1998mq}%
  \BibitemOpen
  \bibfield  {author} {\bibinfo {author} {\bibfnamefont {A.}~\bibnamefont
  {Lue}}, \bibinfo {author} {\bibfnamefont {L.-M.}\ \bibnamefont {Wang}}, \
  and\ \bibinfo {author} {\bibfnamefont {M.}~\bibnamefont {Kamionkowski}},\
  }\href {\doibase 10.1103/PhysRevLett.83.1506} {\bibfield  {journal} {\bibinfo
   {journal} {Phys. Rev. Lett.}\ }\textbf {\bibinfo {volume} {83}},\ \bibinfo
  {pages} {1506} (\bibinfo {year} {1999})},\ \Eprint
  {http://arxiv.org/abs/astro-ph/9812088} {arXiv:astro-ph/9812088 [astro-ph]}
  \BibitemShut {NoStop}%
%%CITATION = ASTRO-PH/9812088;%%
\bibitem [{\citenamefont {Lyth}\ \emph {et~al.}(2005)\citenamefont {Lyth},
  \citenamefont {Quimbay},\ and\ \citenamefont {Rodriguez}}]{Lyth:2005jf}%
  \BibitemOpen
  \bibfield  {author} {\bibinfo {author} {\bibfnamefont {D.~H.}\ \bibnamefont
  {Lyth}}, \bibinfo {author} {\bibfnamefont {C.}~\bibnamefont {Quimbay}}, \
  and\ \bibinfo {author} {\bibfnamefont {Y.}~\bibnamefont {Rodriguez}},\ }\href
  {\doibase 10.1088/1126-6708/2005/03/016} {\bibfield  {journal} {\bibinfo
  {journal} {JHEP}\ }\textbf {\bibinfo {volume} {03}},\ \bibinfo {pages} {016}
  (\bibinfo {year} {2005})},\ \Eprint {http://arxiv.org/abs/hep-th/0501153}
  {arXiv:hep-th/0501153 [hep-th]} \BibitemShut {NoStop}%
%%CITATION = HEP-TH/0501153;%%
\bibitem [{\citenamefont {Satoh}\ \emph {et~al.}(2008)\citenamefont {Satoh},
  \citenamefont {Kanno},\ and\ \citenamefont {Soda}}]{Satoh:2007gn}%
  \BibitemOpen
  \bibfield  {author} {\bibinfo {author} {\bibfnamefont {M.}~\bibnamefont
  {Satoh}}, \bibinfo {author} {\bibfnamefont {S.}~\bibnamefont {Kanno}}, \ and\
  \bibinfo {author} {\bibfnamefont {J.}~\bibnamefont {Soda}},\ }\href {\doibase
  10.1103/PhysRevD.77.023526} {\bibfield  {journal} {\bibinfo  {journal} {Phys.
  Rev.}\ }\textbf {\bibinfo {volume} {D77}},\ \bibinfo {pages} {023526}
  (\bibinfo {year} {2008})},\ \Eprint {http://arxiv.org/abs/0706.3585}
  {arXiv:0706.3585 [astro-ph]} \BibitemShut {NoStop}%
%%CITATION = ARXIV:0706.3585;%%
\bibitem [{\citenamefont {Garretson}\ \emph {et~al.}(1992)\citenamefont
  {Garretson}, \citenamefont {Field},\ and\ \citenamefont
  {Carroll}}]{Garretson:1992vt}%
  \BibitemOpen
  \bibfield  {author} {\bibinfo {author} {\bibfnamefont {W.~D.}\ \bibnamefont
  {Garretson}}, \bibinfo {author} {\bibfnamefont {G.~B.}\ \bibnamefont
  {Field}}, \ and\ \bibinfo {author} {\bibfnamefont {S.~M.}\ \bibnamefont
  {Carroll}},\ }\href {\doibase 10.1103/PhysRevD.46.5346} {\bibfield  {journal}
  {\bibinfo  {journal} {Phys.Rev.}\ }\textbf {\bibinfo {volume} {D46}},\
  \bibinfo {pages} {5346} (\bibinfo {year} {1992})},\ \Eprint
  {http://arxiv.org/abs/hep-ph/9209238} {arXiv:hep-ph/9209238 [hep-ph]}
  \BibitemShut {NoStop}%
%%CITATION = HEP-PH/9209238;%%
\bibitem [{\citenamefont {Anber}\ and\ \citenamefont
  {Sorbo}(2006)}]{Anber:2006xt}%
  \BibitemOpen
  \bibfield  {author} {\bibinfo {author} {\bibfnamefont {M.~M.}\ \bibnamefont
  {Anber}}\ and\ \bibinfo {author} {\bibfnamefont {L.}~\bibnamefont {Sorbo}},\
  }\href {\doibase 10.1088/1475-7516/2006/10/018} {\bibfield  {journal}
  {\bibinfo  {journal} {JCAP}\ }\textbf {\bibinfo {volume} {0610}},\ \bibinfo
  {pages} {018} (\bibinfo {year} {2006})},\ \Eprint
  {http://arxiv.org/abs/astro-ph/0606534} {arXiv:astro-ph/0606534 [astro-ph]}
  \BibitemShut {NoStop}%
%%CITATION = ASTRO-PH/0606534;%%
\bibitem [{\citenamefont {Caprini}\ and\ \citenamefont
  {Sorbo}(2014)}]{Caprini:2014mja}%
  \BibitemOpen
  \bibfield  {author} {\bibinfo {author} {\bibfnamefont {C.}~\bibnamefont
  {Caprini}}\ and\ \bibinfo {author} {\bibfnamefont {L.}~\bibnamefont
  {Sorbo}},\ }\href {\doibase 10.1088/1475-7516/2014/10/056} {\bibfield
  {journal} {\bibinfo  {journal} {JCAP}\ }\textbf {\bibinfo {volume} {1410}},\
  \bibinfo {pages} {056} (\bibinfo {year} {2014})},\ \Eprint
  {http://arxiv.org/abs/1407.2809} {arXiv:1407.2809 [astro-ph.CO]} \BibitemShut
  {NoStop}%
%%CITATION = ARXIV:1407.2809;%%
\bibitem [{\citenamefont {Sorbo}(2011)}]{Sorbo:2011rz}%
  \BibitemOpen
  \bibfield  {author} {\bibinfo {author} {\bibfnamefont {L.}~\bibnamefont
  {Sorbo}},\ }\href {\doibase 10.1088/1475-7516/2011/06/003} {\bibfield
  {journal} {\bibinfo  {journal} {JCAP}\ }\textbf {\bibinfo {volume} {1106}},\
  \bibinfo {pages} {003} (\bibinfo {year} {2011})},\ \Eprint
  {http://arxiv.org/abs/1101.1525} {arXiv:1101.1525 [astro-ph.CO]} \BibitemShut
  {NoStop}%
%%CITATION = ARXIV:1101.1525;%%
\bibitem [{\citenamefont {Anber}\ and\ \citenamefont
  {Sorbo}(2012)}]{Anber:2012du}%
  \BibitemOpen
  \bibfield  {author} {\bibinfo {author} {\bibfnamefont {M.~M.}\ \bibnamefont
  {Anber}}\ and\ \bibinfo {author} {\bibfnamefont {L.}~\bibnamefont {Sorbo}},\
  }\href {\doibase 10.1103/PhysRevD.85.123537} {\bibfield  {journal} {\bibinfo
  {journal} {Phys. Rev.}\ }\textbf {\bibinfo {volume} {D85}},\ \bibinfo {pages}
  {123537} (\bibinfo {year} {2012})},\ \Eprint {http://arxiv.org/abs/1203.5849}
  {arXiv:1203.5849 [astro-ph.CO]} \BibitemShut {NoStop}%
%%CITATION = ARXIV:1203.5849;%%
\bibitem [{\citenamefont {Adshead}\ \emph
  {et~al.}(2013{\natexlab{a}})\citenamefont {Adshead}, \citenamefont
  {Martinec},\ and\ \citenamefont {Wyman}}]{Adshead:2013qp}%
  \BibitemOpen
  \bibfield  {author} {\bibinfo {author} {\bibfnamefont {P.}~\bibnamefont
  {Adshead}}, \bibinfo {author} {\bibfnamefont {E.}~\bibnamefont {Martinec}}, \
  and\ \bibinfo {author} {\bibfnamefont {M.}~\bibnamefont {Wyman}},\ }\href
  {\doibase 10.1103/PhysRevD.88.021302} {\bibfield  {journal} {\bibinfo
  {journal} {Phys. Rev.}\ }\textbf {\bibinfo {volume} {D88}},\ \bibinfo {pages}
  {021302} (\bibinfo {year} {2013}{\natexlab{a}})},\ \Eprint
  {http://arxiv.org/abs/1301.2598} {arXiv:1301.2598 [hep-th]} \BibitemShut
  {NoStop}%
%%CITATION = ARXIV:1301.2598;%%
\bibitem [{\citenamefont {Adshead}\ \emph
  {et~al.}(2013{\natexlab{b}})\citenamefont {Adshead}, \citenamefont
  {Martinec},\ and\ \citenamefont {Wyman}}]{Adshead:2013nka}%
  \BibitemOpen
  \bibfield  {author} {\bibinfo {author} {\bibfnamefont {P.}~\bibnamefont
  {Adshead}}, \bibinfo {author} {\bibfnamefont {E.}~\bibnamefont {Martinec}}, \
  and\ \bibinfo {author} {\bibfnamefont {M.}~\bibnamefont {Wyman}},\ }\href
  {\doibase 10.1007/JHEP09(2013)087} {\bibfield  {journal} {\bibinfo  {journal}
  {JHEP}\ }\textbf {\bibinfo {volume} {09}},\ \bibinfo {pages} {087} (\bibinfo
  {year} {2013}{\natexlab{b}})},\ \Eprint {http://arxiv.org/abs/1305.2930}
  {arXiv:1305.2930 [hep-th]} \BibitemShut {NoStop}%
%%CITATION = ARXIV:1305.2930;%%
\bibitem [{\citenamefont {Maleknejad}(2014)}]{Maleknejad:2014wsa}%
  \BibitemOpen
  \bibfield  {author} {\bibinfo {author} {\bibfnamefont {A.}~\bibnamefont
  {Maleknejad}},\ }\href {\doibase 10.1103/PhysRevD.90.023542} {\bibfield
  {journal} {\bibinfo  {journal} {Phys. Rev.}\ }\textbf {\bibinfo {volume}
  {D90}},\ \bibinfo {pages} {023542} (\bibinfo {year} {2014})},\ \Eprint
  {http://arxiv.org/abs/1401.7628} {arXiv:1401.7628 [hep-th]} \BibitemShut
  {NoStop}%
%%CITATION = ARXIV:1401.7628;%%
\bibitem [{\citenamefont {Maleknejad}(2016)}]{Maleknejad:2016qjz}%
  \BibitemOpen
  \bibfield  {author} {\bibinfo {author} {\bibfnamefont {A.}~\bibnamefont
  {Maleknejad}},\ }\href {\doibase 10.1007/JHEP07(2016)104} {\bibfield
  {journal} {\bibinfo  {journal} {JHEP}\ }\textbf {\bibinfo {volume} {07}},\
  \bibinfo {pages} {104} (\bibinfo {year} {2016})},\ \Eprint
  {http://arxiv.org/abs/1604.03327} {arXiv:1604.03327 [hep-ph]} \BibitemShut
  {NoStop}%
%%CITATION = ARXIV:1604.03327;%%
\bibitem [{\citenamefont {Obata}\ and\ \citenamefont
  {Soda}(2016{\natexlab{a}})}]{Obata:2016tmo}%
  \BibitemOpen
  \bibfield  {author} {\bibinfo {author} {\bibfnamefont {I.}~\bibnamefont
  {Obata}}\ and\ \bibinfo {author} {\bibfnamefont {J.}~\bibnamefont {Soda}}
  (\bibinfo {collaboration} {CLEO}),\ }\href {\doibase
  10.1103/PhysRevD.93.123502} {\bibfield  {journal} {\bibinfo  {journal} {Phys.
  Rev.}\ }\textbf {\bibinfo {volume} {D93}},\ \bibinfo {pages} {123502}
  (\bibinfo {year} {2016}{\natexlab{a}})},\ \Eprint
  {http://arxiv.org/abs/1602.06024} {arXiv:1602.06024 [hep-th]} \BibitemShut
  {NoStop}%
%%CITATION = ARXIV:1602.06024;%%
\bibitem [{\citenamefont {Obata}\ and\ \citenamefont
  {Soda}(2016{\natexlab{b}})}]{Obata:2016xcr}%
  \BibitemOpen
  \bibfield  {author} {\bibinfo {author} {\bibfnamefont {I.}~\bibnamefont
  {Obata}}\ and\ \bibinfo {author} {\bibfnamefont {J.}~\bibnamefont {Soda}},\
  }\href {\doibase 10.1103/PhysRevD.94.044062} {\bibfield  {journal} {\bibinfo
  {journal} {Phys. Rev.}\ }\textbf {\bibinfo {volume} {D94}},\ \bibinfo {pages}
  {044062} (\bibinfo {year} {2016}{\natexlab{b}})},\ \Eprint
  {http://arxiv.org/abs/1607.01847} {arXiv:1607.01847 [astro-ph.CO]}
  \BibitemShut {NoStop}%
%%CITATION = ARXIV:1607.01847;%%
\bibitem [{\citenamefont {Takahashi}\ and\ \citenamefont
  {Soda}(2009)}]{Takahashi:2009wc}%
  \BibitemOpen
  \bibfield  {author} {\bibinfo {author} {\bibfnamefont {T.}~\bibnamefont
  {Takahashi}}\ and\ \bibinfo {author} {\bibfnamefont {J.}~\bibnamefont
  {Soda}},\ }\href {\doibase 10.1103/PhysRevLett.102.231301} {\bibfield
  {journal} {\bibinfo  {journal} {Phys. Rev. Lett.}\ }\textbf {\bibinfo
  {volume} {102}},\ \bibinfo {pages} {231301} (\bibinfo {year} {2009})},\
  \Eprint {http://arxiv.org/abs/0904.0554} {arXiv:0904.0554 [hep-th]}
  \BibitemShut {NoStop}%
%%CITATION = ARXIV:0904.0554;%%
\bibitem [{\citenamefont {Enqvist}\ \emph {et~al.}(2012)\citenamefont
  {Enqvist}, \citenamefont {Figueroa},\ and\ \citenamefont
  {Meriniemi}}]{Enqvist:2012im}%
  \BibitemOpen
  \bibfield  {author} {\bibinfo {author} {\bibfnamefont {K.}~\bibnamefont
  {Enqvist}}, \bibinfo {author} {\bibfnamefont {D.~G.}\ \bibnamefont
  {Figueroa}}, \ and\ \bibinfo {author} {\bibfnamefont {T.}~\bibnamefont
  {Meriniemi}},\ }\href {\doibase 10.1103/PhysRevD.86.061301} {\bibfield
  {journal} {\bibinfo  {journal} {Phys. Rev.}\ }\textbf {\bibinfo {volume}
  {D86}},\ \bibinfo {pages} {061301} (\bibinfo {year} {2012})},\ \Eprint
  {http://arxiv.org/abs/1203.4943} {arXiv:1203.4943 [astro-ph.CO]} \BibitemShut
  {NoStop}%
%%CITATION = ARXIV:1203.4943;%%
\bibitem [{\citenamefont {Barnaby}\ \emph
  {et~al.}(2012{\natexlab{a}})\citenamefont {Barnaby}, \citenamefont {Moxon},
  \citenamefont {Namba}, \citenamefont {Peloso}, \citenamefont {Shiu},\ and\
  \citenamefont {Zhou}}]{Barnaby:2012xt}%
  \BibitemOpen
  \bibfield  {author} {\bibinfo {author} {\bibfnamefont {N.}~\bibnamefont
  {Barnaby}}, \bibinfo {author} {\bibfnamefont {J.}~\bibnamefont {Moxon}},
  \bibinfo {author} {\bibfnamefont {R.}~\bibnamefont {Namba}}, \bibinfo
  {author} {\bibfnamefont {M.}~\bibnamefont {Peloso}}, \bibinfo {author}
  {\bibfnamefont {G.}~\bibnamefont {Shiu}}, \ and\ \bibinfo {author}
  {\bibfnamefont {P.}~\bibnamefont {Zhou}},\ }\href {\doibase
  10.1103/PhysRevD.86.103508} {\bibfield  {journal} {\bibinfo  {journal} {Phys.
  Rev.}\ }\textbf {\bibinfo {volume} {D86}},\ \bibinfo {pages} {103508}
  (\bibinfo {year} {2012}{\natexlab{a}})},\ \Eprint
  {http://arxiv.org/abs/1206.6117} {arXiv:1206.6117 [astro-ph.CO]} \BibitemShut
  {NoStop}%
%%CITATION = ARXIV:1206.6117;%%
\bibitem [{\citenamefont {Figueroa}(2014)}]{Figueroa:2014aya}%
  \BibitemOpen
  \bibfield  {author} {\bibinfo {author} {\bibfnamefont {D.~G.}\ \bibnamefont
  {Figueroa}},\ }\href {\doibase 10.1007/JHEP11(2014)145} {\bibfield  {journal}
  {\bibinfo  {journal} {JHEP}\ }\textbf {\bibinfo {volume} {11}},\ \bibinfo
  {pages} {145} (\bibinfo {year} {2014})},\ \Eprint
  {http://arxiv.org/abs/1402.1345} {arXiv:1402.1345 [astro-ph.CO]} \BibitemShut
  {NoStop}%
%%CITATION = ARXIV:1402.1345;%%
\bibitem [{\citenamefont {Figueroa}\ and\ \citenamefont
  {Meriniemi}(2013)}]{Figueroa:2013vif}%
  \BibitemOpen
  \bibfield  {author} {\bibinfo {author} {\bibfnamefont {D.~G.}\ \bibnamefont
  {Figueroa}}\ and\ \bibinfo {author} {\bibfnamefont {T.}~\bibnamefont
  {Meriniemi}},\ }\href {\doibase 10.1007/JHEP10(2013)101} {\bibfield
  {journal} {\bibinfo  {journal} {JHEP}\ }\textbf {\bibinfo {volume} {10}},\
  \bibinfo {pages} {101} (\bibinfo {year} {2013})},\ \Eprint
  {http://arxiv.org/abs/1306.6911} {arXiv:1306.6911 [astro-ph.CO]} \BibitemShut
  {NoStop}%
%%CITATION = ARXIV:1306.6911;%%
\bibitem [{\citenamefont {Landete}\ \emph {et~al.}(2013)\citenamefont
  {Landete}, \citenamefont {Navarro-Salas},\ and\ \citenamefont
  {Torrenti}}]{Landete:2013axa}%
  \BibitemOpen
  \bibfield  {author} {\bibinfo {author} {\bibfnamefont {A.}~\bibnamefont
  {Landete}}, \bibinfo {author} {\bibfnamefont {J.}~\bibnamefont
  {Navarro-Salas}}, \ and\ \bibinfo {author} {\bibfnamefont {F.}~\bibnamefont
  {Torrenti}},\ }\href {\doibase 10.1103/PhysRevD.88.061501} {\bibfield
  {journal} {\bibinfo  {journal} {Phys. Rev.}\ }\textbf {\bibinfo {volume}
  {D88}},\ \bibinfo {pages} {061501} (\bibinfo {year} {2013})},\ \Eprint
  {http://arxiv.org/abs/1305.7374} {arXiv:1305.7374 [gr-qc]} \BibitemShut
  {NoStop}%
%%CITATION = ARXIV:1305.7374;%%
\bibitem [{\citenamefont {Landete}\ \emph {et~al.}(2014)\citenamefont
  {Landete}, \citenamefont {Navarro-Salas},\ and\ \citenamefont
  {Torrenti}}]{Landete:2013lpa}%
  \BibitemOpen
  \bibfield  {author} {\bibinfo {author} {\bibfnamefont {A.}~\bibnamefont
  {Landete}}, \bibinfo {author} {\bibfnamefont {J.}~\bibnamefont
  {Navarro-Salas}}, \ and\ \bibinfo {author} {\bibfnamefont {F.}~\bibnamefont
  {Torrenti}},\ }\href {\doibase 10.1103/PhysRevD.89.044030} {\bibfield
  {journal} {\bibinfo  {journal} {Phys. Rev.}\ }\textbf {\bibinfo {volume}
  {D89}},\ \bibinfo {pages} {044030} (\bibinfo {year} {2014})},\ \Eprint
  {http://arxiv.org/abs/1311.4958} {arXiv:1311.4958 [gr-qc]} \BibitemShut
  {NoStop}%
%%CITATION = ARXIV:1311.4958;%%
\bibitem [{\citenamefont {del Rio}\ \emph {et~al.}(2017)\citenamefont {del
  Rio}, \citenamefont {Ferreiro}, \citenamefont {Navarro-Salas},\ and\
  \citenamefont {Torrenti}}]{delRio:2017iib}%
  \BibitemOpen
  \bibfield  {author} {\bibinfo {author} {\bibfnamefont {A.}~\bibnamefont {del
  Rio}}, \bibinfo {author} {\bibfnamefont {A.}~\bibnamefont {Ferreiro}},
  \bibinfo {author} {\bibfnamefont {J.}~\bibnamefont {Navarro-Salas}}, \ and\
  \bibinfo {author} {\bibfnamefont {F.}~\bibnamefont {Torrenti}},\ }\href
  {\doibase 10.1103/PhysRevD.95.105003} {\bibfield  {journal} {\bibinfo
  {journal} {Phys. Rev.}\ }\textbf {\bibinfo {volume} {D95}},\ \bibinfo {pages}
  {105003} (\bibinfo {year} {2017})},\ \Eprint
  {http://arxiv.org/abs/1703.00908} {arXiv:1703.00908 [gr-qc]} \BibitemShut
  {NoStop}%
%%CITATION = ARXIV:1703.00908;%%
\bibitem [{\citenamefont {Greene}\ and\ \citenamefont
  {Kofman}(1999)}]{Greene:1998nh}%
  \BibitemOpen
  \bibfield  {author} {\bibinfo {author} {\bibfnamefont {P.~B.}\ \bibnamefont
  {Greene}}\ and\ \bibinfo {author} {\bibfnamefont {L.}~\bibnamefont
  {Kofman}},\ }\href {\doibase 10.1016/S0370-2693(99)00020-9} {\bibfield
  {journal} {\bibinfo  {journal} {Phys. Lett.}\ }\textbf {\bibinfo {volume}
  {B448}},\ \bibinfo {pages} {6} (\bibinfo {year} {1999})},\ \Eprint
  {http://arxiv.org/abs/hep-ph/9807339} {arXiv:hep-ph/9807339 [hep-ph]}
  \BibitemShut {NoStop}%
%%CITATION = HEP-PH/9807339;%%
\bibitem [{\citenamefont {Giudice}\ \emph {et~al.}(1999)\citenamefont
  {Giudice}, \citenamefont {Peloso}, \citenamefont {Riotto},\ and\
  \citenamefont {Tkachev}}]{Giudice:1999fb}%
  \BibitemOpen
  \bibfield  {author} {\bibinfo {author} {\bibfnamefont {G.~F.}\ \bibnamefont
  {Giudice}}, \bibinfo {author} {\bibfnamefont {M.}~\bibnamefont {Peloso}},
  \bibinfo {author} {\bibfnamefont {A.}~\bibnamefont {Riotto}}, \ and\ \bibinfo
  {author} {\bibfnamefont {I.}~\bibnamefont {Tkachev}},\ }\href {\doibase
  10.1088/1126-6708/1999/08/014} {\bibfield  {journal} {\bibinfo  {journal}
  {JHEP}\ }\textbf {\bibinfo {volume} {08}},\ \bibinfo {pages} {014} (\bibinfo
  {year} {1999})},\ \Eprint {http://arxiv.org/abs/hep-ph/9905242}
  {arXiv:hep-ph/9905242 [hep-ph]} \BibitemShut {NoStop}%
%%CITATION = HEP-PH/9905242;%%
\bibitem [{\citenamefont {Greene}\ and\ \citenamefont
  {Kofman}(2000)}]{Greene:2000ew}%
  \BibitemOpen
  \bibfield  {author} {\bibinfo {author} {\bibfnamefont {P.~B.}\ \bibnamefont
  {Greene}}\ and\ \bibinfo {author} {\bibfnamefont {L.}~\bibnamefont
  {Kofman}},\ }\href {\doibase 10.1103/PhysRevD.62.123516} {\bibfield
  {journal} {\bibinfo  {journal} {Phys. Rev.}\ }\textbf {\bibinfo {volume}
  {D62}},\ \bibinfo {pages} {123516} (\bibinfo {year} {2000})},\ \Eprint
  {http://arxiv.org/abs/hep-ph/0003018} {arXiv:hep-ph/0003018 [hep-ph]}
  \BibitemShut {NoStop}%
%%CITATION = HEP-PH/0003018;%%
\bibitem [{\citenamefont {Peloso}\ and\ \citenamefont
  {Sorbo}(2000)}]{Peloso:2000hy}%
  \BibitemOpen
  \bibfield  {author} {\bibinfo {author} {\bibfnamefont {M.}~\bibnamefont
  {Peloso}}\ and\ \bibinfo {author} {\bibfnamefont {L.}~\bibnamefont {Sorbo}},\
  }\href {\doibase 10.1088/1126-6708/2000/05/016} {\bibfield  {journal}
  {\bibinfo  {journal} {JHEP}\ }\textbf {\bibinfo {volume} {05}},\ \bibinfo
  {pages} {016} (\bibinfo {year} {2000})},\ \Eprint
  {http://arxiv.org/abs/hep-ph/0003045} {arXiv:hep-ph/0003045 [hep-ph]}
  \BibitemShut {NoStop}%
%%CITATION = HEP-PH/0003045;%%
\bibitem [{\citenamefont {Berges}\ \emph {et~al.}(2011)\citenamefont {Berges},
  \citenamefont {Gelfand},\ and\ \citenamefont {Pruschke}}]{Berges:2010zv}%
  \BibitemOpen
  \bibfield  {author} {\bibinfo {author} {\bibfnamefont {J.}~\bibnamefont
  {Berges}}, \bibinfo {author} {\bibfnamefont {D.}~\bibnamefont {Gelfand}}, \
  and\ \bibinfo {author} {\bibfnamefont {J.}~\bibnamefont {Pruschke}},\ }\href
  {\doibase 10.1103/PhysRevLett.107.061301} {\bibfield  {journal} {\bibinfo
  {journal} {Phys. Rev. Lett.}\ }\textbf {\bibinfo {volume} {107}},\ \bibinfo
  {pages} {061301} (\bibinfo {year} {2011})},\ \Eprint
  {http://arxiv.org/abs/1012.4632} {arXiv:1012.4632 [hep-ph]} \BibitemShut
  {NoStop}%
%%CITATION = ARXIV:1012.4632;%%
\bibitem [{\citenamefont {Adshead}\ and\ \citenamefont
  {Sfakianakis}(2015)}]{Adshead:2015kza}%
  \BibitemOpen
  \bibfield  {author} {\bibinfo {author} {\bibfnamefont {P.}~\bibnamefont
  {Adshead}}\ and\ \bibinfo {author} {\bibfnamefont {E.~I.}\ \bibnamefont
  {Sfakianakis}},\ }\href {\doibase 10.1088/1475-7516/2015/11/021} {\bibfield
  {journal} {\bibinfo  {journal} {JCAP}\ }\textbf {\bibinfo {volume} {1511}},\
  \bibinfo {pages} {021} (\bibinfo {year} {2015})},\ \Eprint
  {http://arxiv.org/abs/1508.00891} {arXiv:1508.00891 [hep-ph]} \BibitemShut
  {NoStop}%
%%CITATION = ARXIV:1508.00891;%%
\bibitem [{\citenamefont {Freese}\ \emph {et~al.}(1990)\citenamefont {Freese},
  \citenamefont {Frieman},\ and\ \citenamefont {Olinto}}]{Freese:1990rb}%
  \BibitemOpen
  \bibfield  {author} {\bibinfo {author} {\bibfnamefont {K.}~\bibnamefont
  {Freese}}, \bibinfo {author} {\bibfnamefont {J.~A.}\ \bibnamefont {Frieman}},
  \ and\ \bibinfo {author} {\bibfnamefont {A.~V.}\ \bibnamefont {Olinto}},\
  }\href {\doibase 10.1103/PhysRevLett.65.3233} {\bibfield  {journal} {\bibinfo
   {journal} {Phys.Rev.Lett.}\ }\textbf {\bibinfo {volume} {65}},\ \bibinfo
  {pages} {3233} (\bibinfo {year} {1990})}\BibitemShut {NoStop}%
%%CITATION = PRLTA,65,3233;%%
\bibitem [{\citenamefont {Kim}\ and\ \citenamefont
  {Carosi}(2010)}]{Kim:2008hd}%
  \BibitemOpen
  \bibfield  {author} {\bibinfo {author} {\bibfnamefont {J.~E.}\ \bibnamefont
  {Kim}}\ and\ \bibinfo {author} {\bibfnamefont {G.}~\bibnamefont {Carosi}},\
  }\href {\doibase 10.1103/RevModPhys.82.557} {\bibfield  {journal} {\bibinfo
  {journal} {Rev. Mod. Phys.}\ }\textbf {\bibinfo {volume} {82}},\ \bibinfo
  {pages} {557} (\bibinfo {year} {2010})},\ \Eprint
  {http://arxiv.org/abs/0807.3125} {arXiv:0807.3125 [hep-ph]} \BibitemShut
  {NoStop}%
%%CITATION = ARXIV:0807.3125;%%
\bibitem [{\citenamefont {Anber}\ and\ \citenamefont
  {Sorbo}(2010)}]{Anber:2009ua}%
  \BibitemOpen
  \bibfield  {author} {\bibinfo {author} {\bibfnamefont {M.~M.}\ \bibnamefont
  {Anber}}\ and\ \bibinfo {author} {\bibfnamefont {L.}~\bibnamefont {Sorbo}},\
  }\href {\doibase 10.1103/PhysRevD.81.043534} {\bibfield  {journal} {\bibinfo
  {journal} {Phys. Rev.}\ }\textbf {\bibinfo {volume} {D81}},\ \bibinfo {pages}
  {043534} (\bibinfo {year} {2010})},\ \Eprint {http://arxiv.org/abs/0908.4089}
  {arXiv:0908.4089 [hep-th]} \BibitemShut {NoStop}%
%%CITATION = ARXIV:0908.4089;%%
\bibitem [{\citenamefont {Barnaby}\ \emph
  {et~al.}(2012{\natexlab{b}})\citenamefont {Barnaby}, \citenamefont {Pajer},\
  and\ \citenamefont {Peloso}}]{Barnaby:2011qe}%
  \BibitemOpen
  \bibfield  {author} {\bibinfo {author} {\bibfnamefont {N.}~\bibnamefont
  {Barnaby}}, \bibinfo {author} {\bibfnamefont {E.}~\bibnamefont {Pajer}}, \
  and\ \bibinfo {author} {\bibfnamefont {M.}~\bibnamefont {Peloso}},\ }\href
  {\doibase 10.1103/PhysRevD.85.023525} {\bibfield  {journal} {\bibinfo
  {journal} {Phys. Rev.}\ }\textbf {\bibinfo {volume} {D85}},\ \bibinfo {pages}
  {023525} (\bibinfo {year} {2012}{\natexlab{b}})},\ \Eprint
  {http://arxiv.org/abs/1110.3327} {arXiv:1110.3327 [astro-ph.CO]} \BibitemShut
  {NoStop}%
%%CITATION = ARXIV:1110.3327;%%
\bibitem [{\citenamefont {Crowder}\ \emph {et~al.}(2013)\citenamefont
  {Crowder}, \citenamefont {Namba}, \citenamefont {Mandic}, \citenamefont
  {Mukohyama},\ and\ \citenamefont {Peloso}}]{Crowder:2012ik}%
  \BibitemOpen
  \bibfield  {author} {\bibinfo {author} {\bibfnamefont {S.~G.}\ \bibnamefont
  {Crowder}}, \bibinfo {author} {\bibfnamefont {R.}~\bibnamefont {Namba}},
  \bibinfo {author} {\bibfnamefont {V.}~\bibnamefont {Mandic}}, \bibinfo
  {author} {\bibfnamefont {S.}~\bibnamefont {Mukohyama}}, \ and\ \bibinfo
  {author} {\bibfnamefont {M.}~\bibnamefont {Peloso}},\ }\href {\doibase
  10.1016/j.physletb.2013.08.077} {\bibfield  {journal} {\bibinfo  {journal}
  {Phys. Lett.}\ }\textbf {\bibinfo {volume} {B726}},\ \bibinfo {pages} {66}
  (\bibinfo {year} {2013})},\ \Eprint {http://arxiv.org/abs/1212.4165}
  {arXiv:1212.4165 [astro-ph.CO]} \BibitemShut {NoStop}%
%%CITATION = ARXIV:1212.4165;%%
\bibitem [{\citenamefont {Bartolo}\ \emph {et~al.}(2016)\citenamefont {Bartolo}
  \emph {et~al.}}]{Bartolo:2016ami}%
  \BibitemOpen
  \bibfield  {author} {\bibinfo {author} {\bibfnamefont {N.}~\bibnamefont
  {Bartolo}} \emph {et~al.},\ }\href {\doibase 10.1088/1475-7516/2016/12/026}
  {\bibfield  {journal} {\bibinfo  {journal} {JCAP}\ }\textbf {\bibinfo
  {volume} {1612}},\ \bibinfo {pages} {026} (\bibinfo {year} {2016})},\ \Eprint
  {http://arxiv.org/abs/1610.06481} {arXiv:1610.06481 [astro-ph.CO]}
  \BibitemShut {NoStop}%
%%CITATION = ARXIV:1610.06481;%%
\bibitem [{\citenamefont {Gerbino}\ \emph {et~al.}(2016)\citenamefont
  {Gerbino}, \citenamefont {Gruppuso}, \citenamefont {Natoli}, \citenamefont
  {Shiraishi},\ and\ \citenamefont {Melchiorri}}]{Gerbino:2016mqb}%
  \BibitemOpen
  \bibfield  {author} {\bibinfo {author} {\bibfnamefont {M.}~\bibnamefont
  {Gerbino}}, \bibinfo {author} {\bibfnamefont {A.}~\bibnamefont {Gruppuso}},
  \bibinfo {author} {\bibfnamefont {P.}~\bibnamefont {Natoli}}, \bibinfo
  {author} {\bibfnamefont {M.}~\bibnamefont {Shiraishi}}, \ and\ \bibinfo
  {author} {\bibfnamefont {A.}~\bibnamefont {Melchiorri}},\ }\href {\doibase
  10.1088/1475-7516/2016/07/044} {\bibfield  {journal} {\bibinfo  {journal}
  {JCAP}\ }\textbf {\bibinfo {volume} {1607}},\ \bibinfo {pages} {044}
  (\bibinfo {year} {2016})},\ \Eprint {http://arxiv.org/abs/1605.09357}
  {arXiv:1605.09357 [astro-ph.CO]} \BibitemShut {NoStop}%
%%CITATION = ARXIV:1605.09357;%%
\bibitem [{\citenamefont {Arkani-Hamed}\ \emph {et~al.}(2016)\citenamefont
  {Arkani-Hamed}, \citenamefont {Cohen}, \citenamefont {D'Agnolo},
  \citenamefont {Hook}, \citenamefont {Kim},\ and\ \citenamefont
  {Pinner}}]{Arkani-Hamed:2016rle}%
  \BibitemOpen
  \bibfield  {author} {\bibinfo {author} {\bibfnamefont {N.}~\bibnamefont
  {Arkani-Hamed}}, \bibinfo {author} {\bibfnamefont {T.}~\bibnamefont {Cohen}},
  \bibinfo {author} {\bibfnamefont {R.~T.}\ \bibnamefont {D'Agnolo}}, \bibinfo
  {author} {\bibfnamefont {A.}~\bibnamefont {Hook}}, \bibinfo {author}
  {\bibfnamefont {H.~D.}\ \bibnamefont {Kim}}, \ and\ \bibinfo {author}
  {\bibfnamefont {D.}~\bibnamefont {Pinner}},\ }\href {\doibase
  10.1103/PhysRevLett.117.251801} {\bibfield  {journal} {\bibinfo  {journal}
  {Phys. Rev. Lett.}\ }\textbf {\bibinfo {volume} {117}},\ \bibinfo {pages}
  {251801} (\bibinfo {year} {2016})},\ \Eprint
  {http://arxiv.org/abs/1607.06821} {arXiv:1607.06821 [hep-ph]} \BibitemShut
  {NoStop}%
%%CITATION = ARXIV:1607.06821;%%
\bibitem [{\citenamefont {Figueroa}\ \emph {et~al.}(2016)\citenamefont
  {Figueroa}, \citenamefont {García-Bellido},\ and\ \citenamefont
  {Torrentí}}]{Figueroa:2016ojl}%
  \BibitemOpen
  \bibfield  {author} {\bibinfo {author} {\bibfnamefont {D.~G.}\ \bibnamefont
  {Figueroa}}, \bibinfo {author} {\bibfnamefont {J.}~\bibnamefont
  {García-Bellido}}, \ and\ \bibinfo {author} {\bibfnamefont {F.}~\bibnamefont
  {Torrentí}},\ }\href {\doibase 10.1103/PhysRevD.93.103521} {\bibfield
  {journal} {\bibinfo  {journal} {Phys. Rev.}\ }\textbf {\bibinfo {volume}
  {D93}},\ \bibinfo {pages} {103521} (\bibinfo {year} {2016})},\ \Eprint
  {http://arxiv.org/abs/1602.03085} {arXiv:1602.03085 [astro-ph.CO]}
  \BibitemShut {NoStop}%
%%CITATION = ARXIV:1602.03085;%%
\end{thebibliography}%

%===================================================================================

\end{document}